\documentclass[A4]{article}

\usepackage{natbib}

\usepackage[utf8]{inputenc}

\usepackage{graphicx}
\usepackage{amsmath, amssymb, setspace}
\usepackage{alltt}
\usepackage{hyperref}
\usepackage[normalem]{ulem}

\usepackage{hyperref}
\usepackage{xcolor}
\definecolor{dark-red}{rgb}{0.4,0.15,0.15}
\definecolor{dark-blue}{rgb}{0.15,0.15,0.8}
\definecolor{medium-blue}{rgb}{0,0,0.5}
\hypersetup{
    colorlinks, linkcolor={dark-red},
    citecolor={dark-blue}, urlcolor={medium-blue}
}

\usepackage{subcaption}
\usepackage{tikz}
\usetikzlibrary{automata,positioning}

\usepackage{amsmath}

\usepackage{tabularx}

\graphicspath{{./figs/}}
\DeclareGraphicsExtensions{.pdf}

\title{Scalability in Computing and Robotics}

\author{Heiko Hamann$^1$ \and Andreagiovanni Reina$^2$}

\date{%
    $^1$ Institute of Computer Engineering, University of L\"ubeck, L\"ubeck, Germany\\%
    $^2$ IRIDIA, Universit\'{e} Libre de Bruxelles, Brussels, Belgium; Department of Computer Science, University of Sheffield, \\Sheffield, UK\\%
    \today
}

\begin{document}


\maketitle

\begin{abstract}
  Efficient engineered systems require scalability. A~scalable system
  has increasing performance with increasing system size. In an ideal
  situation, the increase in performance (\textit{e.g.}, speedup)
  corresponds to the number of units (\textit{e.g.}, processors,
  robots, users) that are added to the system (\textit{e.g.}, three times the
  number of processors in a computer would lead to three times faster
  computations). However, if multiple units work on the same task,
  then coordination among these units is required. This coordination
  can introduce overheads with an impact on system performance. The
  coordination costs can lead to sublinear improvement or even
  diminishing performance with increasing system size. However, there
  are also systems that implement efficient coordination and exploit
  collaboration of units to attain superlinear improvement. Modeling
  the scalability dynamics is key to understanding and engineering
  efficient systems. Known laws of scalability, such as Amdahl's law,
  Gustafson's law, and Gunther's Universal Scalability Law, are
  minimalistic phenomenological models that explain a rich variety of
  system behaviors through concise equations. While useful to gain general
  insights, the phenomenological nature of these models may limit the
  understanding of the underlying dynamics, as they are detached from
  first principles that could explain coordination overheads or
  synergies among units. Through a decentralized system approach, we
  propose a general model based on generic interactions between units
  that is able to describe, as specific cases, any general pattern of
  scalability included by previously reported laws. The proposed
  general model of scalability has the advantage of being built on
  first principles, or at least on a microscopic description of
  interaction between units, and therefore has the potential to
  contribute to a better understanding of system behavior and
  scalability.  We show that this generic model can be applied to a
  diverse set of systems, such as parallel supercomputers, robot
  swarms, or wireless sensor networks, therefore creating a unified
  view on interdisciplinary design for scalability.
\end{abstract}


%

\section{Introduction}
Many engineered systems can improve their performance by
parallelizing the execution of tasks among its constituent
units~\citep{kirk2016programming, hamann2018book}. Increasing the
system size~$N$ (number of units) can lead to an improvement in the
system performance, 
that we understand here as mean throughput~$X(N)$ (completed sub-tasks per time).\footnote{Alternatively, one can define speedup based on latencies: $S=T(1)/T(N)$ with $T(1)$ is the time one has to wait until the single-unit system is done and $T(N)$ is the time of the $N$-unit system.} A~linear speedup of~$S_\text{thruput}(N)=X(N)/X(1)=N$ is observed when 
the $N$-unit system achieves $N$-times the throughput of 
the
single-unit system~($N=1$). 
However, most systems cannot
be scaled up arbitrarily because parallelization comes with a number
of constraints. Some tasks, for instance, may not allow to be fully
parallelized but have a certain fraction~$\sigma$ of necessarily
serial operations. In addition, parallel computing requires a certain
level of coordination among units that may negatively impact on the
overall performance~\citep{archibald1986cache,farkas1995scalable}.
In order to design, predict, and control engineered systems, models describing the system's scalability are needed. 

We propose a new model for system scalability that has increased explanatory power compared to previously published approaches~\citep{amdahl1967validity,gustafson1988,gunther93}. 
First, we abstract the behavior of individual system units to a three-state finite state machine.
That state machine also serves as a population model representing percentages of the system's units that stay in each of these three states during a transient and in equilibrium. 
Second, we link the microscopic level (individual unit behavior) with the macroscopic level (system behavior) as we start from reaction equations of individual behavior and give the respective population model. 
Third, we derive previous scalability models as special cases of ours. 
Besides the formulation of the model, our second main contribution is to point to the generality of system scalability behaviors across diverse fields, such as computing, robotics, and networks.

\subsection{Laws of scalability}

\begin{figure}
    \begin{subfigure}[c]{0.45\textwidth}
      \includegraphics[width=1\textwidth]{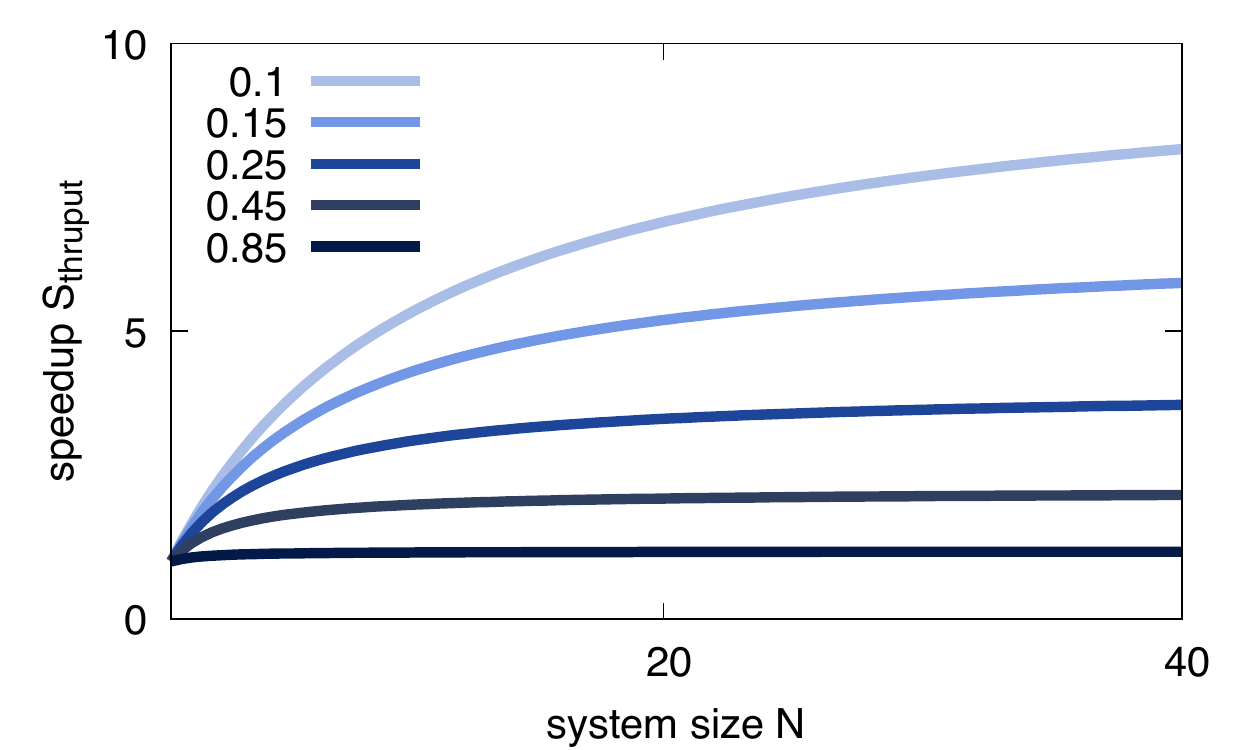}
      \subcaption{\label{fig:amdahl}Amdahl's law}
    \end{subfigure}
    \begin{subfigure}[c]{0.45\textwidth}
      \includegraphics[width=1\textwidth]{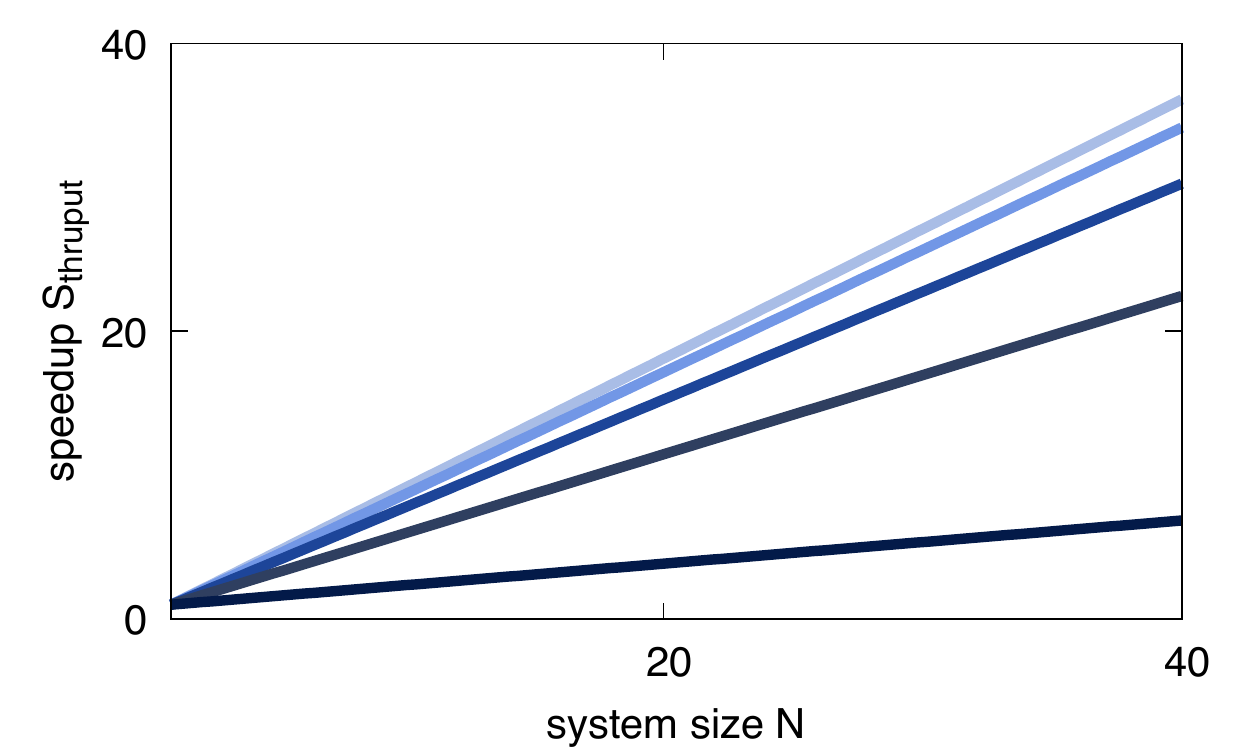}
      \subcaption{\label{fig:gustafson}Gustafson's law}
    \end{subfigure}\\
    \begin{subfigure}[c]{0.45\textwidth}
      \includegraphics[width=1\textwidth]{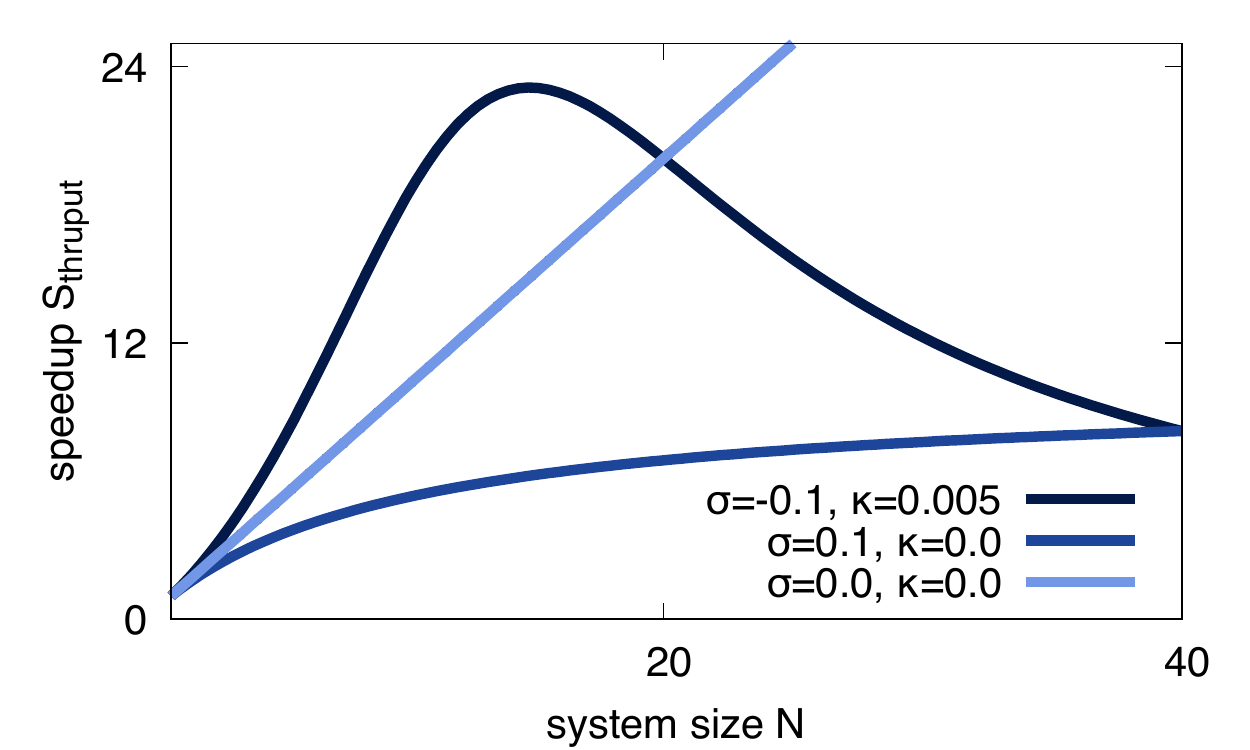}
      \subcaption{\label{fig:gunther}Gunther's USL}
    \end{subfigure}
    \caption{Speedup $S_\text{thruput}$ as a function of the system size~$N$
      (\textit{e.g.}, number of processors, robots, users) according to 
      three known laws of scalability. (\subref{fig:amdahl})~Amdahl's
      law (Eq.~\ref{eq:amdahl}) for serial proportion parameter
      $\sigma\in\{0.1,0.15,0.25,0.45,0.85\}$ shows performance
      increase for small $N$ and saturation to a constant performance
      for large $N$. (\subref{fig:gustafson})~Gustafson's law
      (Eq.~\ref{eq:gustafson}) shows a unbounded increase of
      performance with increasing system
      size. (\subref{fig:gunther})~Gunther's Universal Scalability Law
      (USL, Eq.~\eqref{eq:usl}) can show a richer set of dynamics including diminishing returns (for positive coherence delay $\kappa>0$) and superlinear speedups (for
      negative contention $\sigma<0$); the USL curves in panel~(\subref{fig:gunther}) is for $\sigma=0$ and 
      $\kappa=0$, $\sigma=0.1$ and 
      $\kappa=0$, $\sigma=-0.1$ and 
      $\kappa=0.005$}
    \label{fig:amdahlGunther}
\end{figure}

One of the first attempts of modeling such system behavior
came from Amdahl~\citep{amdahl1967validity} who was motivated by
empirical evidence and aimed to highlight the limitations of parallel
computing. While the actual equation was not provided in the initial
publication, Amdahl's law was later formalized as
a function describing the speedup~
\mbox{$S_\text{thruput}(N)=X(N)/X(1)$}. 
Amdahl's law reads as
\begin{equation}
\label{eq:amdahl}
S_\text{thruput}(N)=\frac{N}{1+\sigma(N-1)}\;\, ,
\end{equation}
where $0\le \sigma \le 1$ is the amount of time spent on serial parts
of the task. As illustrated in Fig.~\ref{fig:amdahl}, Amdahl's law
predicts a bounded scalability of performance, that is, $S_\text{thruput}(N)$ 
saturates at an approximately constant value of $\sigma^{-1}$ for any
large $N$.

Later, a more optimistic perspective was provided by Gustafson's
law~\citep{gustafson1988}
\begin{equation}
\label{eq:gustafson}
S_\text{thruput}(N)=N+(1-N)\sigma\;\, ,
\end{equation}
which assumes that the impact of the serial part $0\le \sigma \le 1$
on the speedup~$S_\text{thruput}(N)$ decreases with system size~$N$. Therefore,
Eq.~\eqref{eq:gustafson} leads to an unbounded growth of~$S_\text{thruput}(N)$ by
increasing~$N$ as shown in Fig.~\ref{fig:gustafson}.


Later, Gunther~\citep{gunther93} drew attention to the possibility that for
large systems, coordination overheads for increased system size
may be larger than the work that can be done by the added units. The
performance of a system with a size larger than critical
value~$N_c$ retrogrades by increasing~$N$. Operating a system with
size~$N>N_c$ is outright undesirable due to the combined cost of
additional units and reduced overall performance.
Instead, it is particularly useful to have models that can predict the
value of $N_c$ and help to avoid operating systems in the region
$N>N_c$. 
Gunther~\citep{gunther93} extended Amdahl's law to define the Universal
Scalability Law (USL) as
\begin{equation}
\label{eq:usl}
 S_\text{thruput}(N)=\frac{N}{1+\sigma(N-1)+\kappa N(N-1)}\;\, , 
\end{equation}
where $\sigma$ is now called contention (time lost queuing for shared resources) and the additional parameter~$\kappa$ accounts for the coherency
delay (overheads of coordinating many units).\footnote{Note that there is an unpublished 3-parameter USL model variant by Gunther where a third parameter~$\gamma$ is multiplied to eq.~\ref{eq:usl}. It is equivalent to a Gustafson factor and can represent the throughput for~$N=1$.} The
underlying idea is that a cost proportional to the square of the
system size~$\propto N^2$ is necessary to administrate the system.
A quadratic increase in the coordination cost corresponds to the
assumption of a coherency delay $\kappa$ caused by the interaction of
every unit with all other~$N-1$ units (all-to-all interaction or a constant fraction of all-to-all interaction resulting in $\mathcal{O}(N^2)$ overhead). This
type of problem is frequent in parallel programs with shared memory
that need to ensure cache coherency, for example, by employing
protocols, such as bus
snooping~\citep{archibald1986cache,farkas1995scalable}.
Note that setting~$\kappa=0$ in Eq.~\eqref{eq:usl} is Amdahl's law (Eq.~\eqref{eq:amdahl}). Setting~$\sigma=0$ and~$\kappa=0$ models a linear speedup~$S_\text{thruput}(N)=N$ similar to Gustafson's law. 
Despite scalability analysis in the region~$N>N_c$ has been largely
ignored by others than Gunther, we argue that knowledge about the system performance in the
full system size domain, including $N>N_c$, can give a better
understanding of the system behavior.

\subsection{Retrograde performance and superlinear speedups}

Retrograding performance can be observed in models in which increasing
the system size triggers a nonlinear slowdown. Therefore, once going beyond the
critical size~$N_c$, the added benefits of additional units are
smaller than the caused slowdown, for example, due to required access to shared
resources. In queuing theory, the \textit{repairman} model
\citep{Allen90} is a simple classical model that describes how a
shared resource can limit the system's throughput. In this model, the
shared resource is the repairman who can only repair machines
sequentially, taking time~$S$ per machine. The model assumes that
machines periodically need downtime (repair time) and shows that for large
number of machines~$N$ the system throughput saturates to~$1/S$~\citep{gunther07}, reproducing the behavior of Amdahl's law. Throughput saturation is caused by each machine getting periodically
unproductive for the repair time~$R$ that sums both the
repairman's service time~$S$ and waiting time~$W$. However, to model retrograde performance, the repair
time~$R$ needs to nonlinearly increase with increasing waiting time~$W$ due to longer queues.
Gunther~\citep{gunther2008general} indicated that assuming a quadratic
increase of~$R$, the repairman model becomes the state-dependent
server model~\textit{M/G/1//p} from queuing theory~\citep{gittins1989}. From this state-dependent model, Gunther~\citep{gunther2008general} was able to derive a specific form of the USL ($\kappa=c\sigma$ for a constant~$c>0$) where the term~$N(N-1)$ in Eq.~\ref{eq:usl} directly follows from nonlinearly increasing repair times~$R\propto N(N-1)$.
In this way, he got close to a scalability law based on first principles but the state-dependent server model already contains the nonlinear repair times as a modeling assumption.
Hence, also this approach has limited potential for insights into the underlying system behavior.
 
Gunther's USL also covers another crucial, and at the same time
controversial, phenomenon of parallel computing: superlinear
speedup, that is, $S_\text{thruput}>N$. Naively, one may assume that linear speedup of~$S_\text{thruput}=N$
is the ideal case. However, Gunther et al.~\citep{gunther15} have shown that
superlinear speedups occur in actual computing systems, such as in
systems for distributed storage and processing of big data
(\textit{e.g.}, Apache Hadoop~\citep{ApacheHadoop}). These surprising
results question the initial doubts about whether superlinear
speedups can exist~\citep{gustafson90,helmbold90}.  In parallel computing,
a common criticism to superlinear speedups has been that a
single-processor machine could, in theory, emulate any $N$-processor
machine to achieve the same speedup performance.  Hence, superlinear
speedup has been claimed to only take place on suboptimal sequential
algorithms~\citep{grama2003introduction}. Such a point of view,
however, ignores physical constraints that could render such an
emulation impossible, such as cache effects or synergistic
collaboration between processors.

Scalability analysis studies the performance for increasing system
size, that is, for adding units. 
A~unit needs to have potential to positively contribute to system throughput.
Since throughput is defined as completed sub-tasks per time, it can be scaled in two dimensions: provided work force and provided sub-tasks. 
A~system can be limited in two ways: shortage in workers and/or shortage in sub-tasks. 
A~unit's contribution, hence, can be of two kinds: adding work force and/or adding sub-tasks.
An added robot in a multi-robot system increases work force and an added user in a database benchmark increases system load.
In both cases, the unit adds to the overall demand on shared resources.
The robot requires a section of shared space to navigate and the user requires computation time on the shared processor.
It is important to understand that
depending on how we define the units, we can obtain different
scalability outcomes. We can assume \textit{pure} units that add
resources in only a single dimension, for example, only a computation module,
or only a memory module. Differently, we can assume \textit{combined}
units that add a 
set of resources, for example, in parallel
computing an added CPU may also include some memory (such as CPU
cache), or in robotics an added robot includes computation, memory,
sensing, and actuation resources. We believe that pure units are
better fitted to theoretical analysis, while combined units are closer
to what is observed in real systems. In parallel computing, adding
computing power without any memory may be unfeasible as CPU cache is
 integrated in microprocessors. Even more so in robotics, adding pure units for computation, for example, would
require redesigning the robots to maintain the memory, sensing, and
actuation of the group constant while increasing the computation
only. In relation to the discussion of superlinearity~\citep{gustafson90,
  helmbold90, grama2003introduction, gunther15, gunther15b}, we believe that the
possibility and impossibility of superlinear speedups is linked to the
distinction between pure and combined units as yet overlooked.
Superlinear speedup is not theoretically possible with pure units,
while it can be observed, both in computing~\citep{gunther15, gunther15b} and in
robotics~\citep{ijspeert01b, ogrady07, Talamali:SwInt:2019},
considering combined units. 
%
In computing, superlinear speedups are a consequence of the combination of increased computing power and cache, in what is a so-called \textit{cache effect}~\citep{gunther15}. Through cache effects, the combined cache of all $N$~CPUs reaches a size large enough to allow the execution of the entire task 
on each CPU without access to external memory. 
In multi-robot systems, superlinear speedups are also caused by a combination of added resources. For example, O'Grady et al.~\citep{ogrady07} showed that a team of $N=3$~physically connected robots can cross a gap that cannot otherwise be bridged by a single robot. In this case, the robots physically connect with each other and hence benefit from a combination of more traction power and a larger base area.
Note that, to observe a superlinear speedup, the workload must be kept constant. If the workload would be increased with the number of CPUs, the cache effect would not be observed. Similarly, if the gap would be made wider for increased robot team size, no team could cross it.

In order to allow the USL to
describe superlinear speedups, parameter~$\sigma$ must be
negative (see Fig.~\ref{fig:gunther}). Therefore, the initial
interpretation of parameter~$\sigma$ as the serial fraction of the
task was changed to interpret it as contention. Positive
contention (\textit{i.e.}, $\sigma>0$) describes the overhead cost due
to coordination of shared resources while negative contention
(\textit{i.e.}, $\sigma<0$) describes the effects of synergistic
interaction among the units, for instance collaborating robots.

\subsection{Scalability in multi-robot systems}

In multi-robot systems, similar performance curves have been observed
and scalability is a key aspect of numerous studies~\citep{hamann2018book, Bjerknes2013, 
  Talamali:SwInt:2019}.  Hamann~\citep{hamann18b} followed a mostly
phenomenological approach by fitting performance curves from the swarm
robotics literature. The adopted function to model the swarm
performance is 
\begin{equation}
\label{eq:swarmPerformance}
  S_\text{thruput}(N)=aN^b\exp(cN)\;\, , 
\end{equation}
for constants $a>0$, $b>0$, and $c<0$. 
The function can be understood
as a dichotomous pair of a term for potential of collaboration~$N^b$
and a term for interference~$\exp(cN)$.
This model was fitted to a number of application scenarios in swarm robotics, such as foraging, collective decision-making, and collective motion~\citep{hamann13a,hamann18b} and it bears similarity to models of the slotted ALOHA communication protocol~\citep{roberts75,gokturk08}.
While the speedup and performance curves in computing and multi-robot systems can be similar, a key difference is that certain multi-robot applications require the collaboration of robots.
For these cases, the task cannot be solved unless two or more robots interact and work together.
Examples are the stick pulling scenario~\citep{ijspeert01b} and the the emergent taxis scenario~\citep{bjerknes07}.
In stick pulling, two robots are required to pull a stick due to mechanical constraints. 
In emergent taxis, robots collectively move towards a beacon  while an individual robot would not be able to do so due to limited sensors.

\subsection{Related efforts in other disciplines}

Our ultimate goal of finding a generic function of performance over system size relates to several efforts in other fields, such as network theory and statistical physics. 
For example, Feng et al.~\citep{feng20} study how the Pareto distribution and the log-normal distribution can be used to fit degree distributions of scale-free networks. 
Similarly to this paper they also validate their approach empirically by fitting their model to data from applications. Note that we focus on functions of speedup over system size but we speculate that the underlying interaction networks (\textit{e.g.}, defined by required coordination of processors or mutual perception of robots) also scale and probably change their node degree distributions significantly~\citep{khaluf17,bettstetter04}. The study by Lazer and Friedman~\citep{lazer07} is an example where group performance is directly related to network topology.

As we try to model interacting computers, robots, and other agents, there are many related fields that have studied similar systems. 
Often a formalism borrowed from statistical physics and chemical kinetics, such as rate equations and other ODE systems, are applied~\citep{espenson95}. 
Rate equations have been used extensively in parallel computing and networks to model stochastic communication, for example,
rumor or virus spreading (gossiping)~\citep{dumitras03,zhang2007performance,bogdan09,sayama13}.

\subsection{Scalability across disciplines}
Models describing the scalability of parallel computers or robot
swarms have been instrumental to gain important insights about the system
dynamics, however they are limited by one common aspect. The models
are phenomenological in the sense that they have been derived to
effectively match the observed throughput~$S_\text{thruput}$ without
explaining how the system \textit{behaves} and therefore why the
performance varies. 
We contribute towards this endeavor by employing our
expertise in decentralized system theory~\citep{resnick94,martinoli04,garnier07,hamann2018book} to formalize a general model
that unifies scalability dynamics of a diversity of systems, from
parallel computers to robot swarms. The proposed model is based
on generic rules that describe interactions between units and how
these interactions change the state of the units and, in turn, of the
whole system. Therefore, the scalability dynamics of our model are the
results of a mechanistic explanation of the system. Such a bottom-up
description paradigm is, for example, core in the emerging field of \textit{machine
  behavior}~\citep{Rahwan2019}. Our results contribute towards the
endeavor to engineer and control machine interactions based on a deep
understanding of their behavior.
%
Thanks to its generality, our model is of potential interest to both
the distributed robotics and the parallel computing communities, as it
can flexibly describe the dynamics of a variety of systems and help in
the understanding of their causes. While we do not discuss it here, we
envision sensor networks, transport systems, chemical systems, and
collective animal behavior to be potential further fields of
application.


\section{Model}

\begin{figure}\centering
\includegraphics[width=1.0\columnwidth]{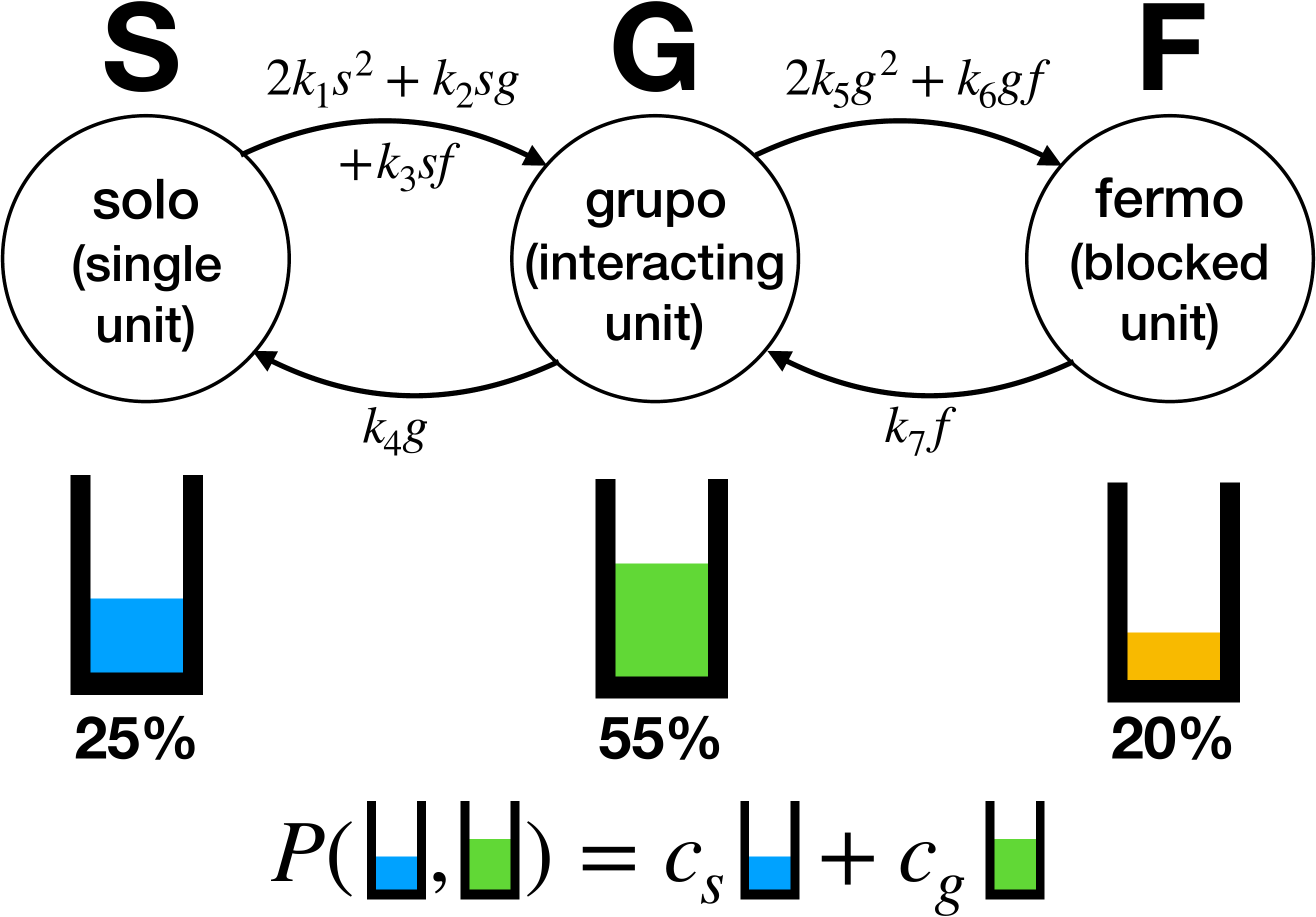}
\caption{
  Graphical representation of the proposed model in the form of a
  three-state machine: solo units (state~$S$), interacting units
  (state~$G$), and blocked units (state~$F$). The number of units is
  constant (Eq.~\eqref{eq:fixed-size}) and they change their state
  (arrows) according to the model of
  Eqs.~\eqref{eq:reaction1}-\eqref{eq:reaction7}. As indicated in
  Eq.~\eqref{eq:performance-function:throughput}, the number of units in each
  state determines the system performance.}
\label{fig:stateMachineAndPerformance}
\end{figure}

Following decentralized system theory, we describe a parallel system
as composed by a number~$N$ of interacting units (\textit{e.g.},
processors, robots, users) that work towards the same task. A~unit at each
moment can be in one of three possible states: \textit{solo}~($S$),
\textit{grupo}~($G$), and \textit{fermo}~($F$).\footnote{These words are loosely adapted from Esperanto and Italian: sole/solo for alone, grupo/gruppo for group, fermo/fermo for stationary or firmly.} 
These states indicate how the unit is
working toward completing the task (see state machine in Fig.~\ref{fig:stateMachineAndPerformance}). A~unit can be either working in
solitary mode~($S$), interacting with other units~($G$), or being
unproductive due to congestion on shared resources~($F$). Depending on
the task and the application, units in each of these states can
contribute in varying degrees to the system performance (as later
indicated by Eq.~\eqref{eq:performance-function:throughput}).

\subsection{Transition rates} 
The model describes changes in a unit's
state via transition rates which determine the conditions and speed of
such changes. The state transitions can be described via reaction
stoichiometry in the form of balanced equations:
\begin{equation}
  \label{eq:reaction1}
  2S \xrightarrow{k_1} 2G
\end{equation}
\begin{equation}
  \label{eq:reaction2}
  S+G \xrightarrow{k_2} 2G
\end{equation}
\begin{equation}
  \label{eq:reaction3}
  S+F \xrightarrow{k_3} G + F
\end{equation}
\begin{equation}
  \label{eq:reaction4}
  G \xrightarrow{k_4} S
\end{equation}
\begin{equation}
  \label{eq:reaction5}
  2G \xrightarrow{k_5} 2F
\end{equation}
\begin{equation}
  \label{eq:reaction6}
  G+F \xrightarrow{k_6} 2F
\end{equation}
\begin{equation}
  \label{eq:reaction7}
  F \xrightarrow{k_7} G \,.
\end{equation}
The seven transitions with respective rates~$k_i$ (with
$i \in \{1,2,\dots,7\}$) determine the system dynamics and can be
interpreted as follows. The transitions of
Eqs.~\eqref{eq:reaction1}-\eqref{eq:reaction3} describe individual
units (state \textit{solo}~$S$) that begin an interaction with other units and
change their state to \textit{grupo}~$G$. The transitions of
Eqs.~\eqref{eq:reaction5} and~\eqref{eq:reaction6} describe interacting
individuals (state \textit{grupo}~$G$) that due to congestion on a shared resource go
to blocked state \textit{fermo}~$F$. Finally, Eqs.~\eqref{eq:reaction4} and
\eqref{eq:reaction7} respectively describe units completing their interactions
and decongesting. Note that units do not directly change state from
\textit{solo}~$S$ to \textit{fermo}~$F$ nor viceversa. Instead, in order to reach or
decongest a \textit{fermo} state, units always go through an interaction with
other units via state \textit{grupo}~$G$ to model cost paid in time due to switching overheads.

The seven transition rates~$k_i$ determine the frequency of transitions as single unit actions
(\textit{i.e.}, Eqs.~\eqref{eq:reaction4} and \eqref{eq:reaction7}) or,
in other cases, conditional to contact between two units in specific
states (\textit{i.e.}, Eqs.~\eqref{eq:reaction1}-\eqref{eq:reaction3},
 \eqref{eq:reaction5} and~\eqref{eq:reaction6}). By modulating the
relative frequency of the seven transition rates~$k_i$, the model can
describe any previously observed scalability law and establish a link
between the units' state change and the whole system performance, as
illustrated in Sec.~\ref{sec:results}.

\subsection{Mathematical model}
Through van Kampen expansion \citep{vankampen}, the model of
Eqs.~\eqref{eq:reaction1}-\eqref{eq:reaction7} can be written in the
form of a system of ordinary differential equations (ODEs):
\begin{equation}
    \left\{
    \begin{aligned}
      \frac{ds}{dt} &=-2k_1 s^2-k_2 s g - k_3 s f + k_4 g \\
      \frac{dg}{dt} &=2 k_1 s^2+ k_2 s g + k_3 s f - k_4 g - 2 k_5 g^2
      - k_6 g f + k_7 f\\
      \frac{df}{dt} &= 2k_5 g^2 + k_6 g f - k_7 f \,,
    \end{aligned}
    \right.
    \label{eq:model:full3ode:maintext}
\end{equation}
where the system variables $s$, $g$, and $f$ represent the number of
units in states \textit{solo}~$S$, \textit{grupo}~$G$, and \textit{fermo}~$F$, respectively.
The model of Eq.~\eqref{eq:model:full3ode:maintext} is a macroscopic
description of the system as it describes how the population of units
is divided between three subpopulations in distinct states and how
these subpopulations change over time.
The proposed model is a closed system that describes a population
comprised of a constant number of units with dynamic states. We can
therefore replace one of the differential equations of
Eq.~\eqref{eq:model:full3ode:maintext} with the equation
\begin{equation}
  N = s + g + f \;,
  \label{eq:fixed-size}
\end{equation}
obtaining a two-equation system (see details in the Appendix).
Through stability analysis, we can compute the fixed points of the
system of Eq.~\eqref{eq:model:full3ode:maintext}. Among these fixed
points, we are interested in studying the stable point
$(s^*, g^*, f^*)$ that exists in the positive simplex, that is,
the three subpopulations are positive ($\ge 0$) and smaller than~$N$ (see
details in the Appendix). Such a stable fixed point can be interpreted
as the final state toward which the system asymptotically converges, that is,
\begin{align}
\label{eq:fixedPoints}
   s^*=\lim_{t\rightarrow \infty}s(t) \,, \;
   g^*=\lim_{t\rightarrow \infty}g(t) \,, \;
   f^*=\lim_{t\rightarrow \infty}f(t) \,,\;
\end{align}
for the starting point $s(0)=N, g(0)=0, f(0)=0$. Studying how the
stable fixed point $(s^*, g^*, f^*)$ changes as a function of the
system size~$N$ (and transition rates) provides information on the system's
scalability. The technical details about the stability analysis are
provided in the Appendix.

Note that we derived the ODE system of Eq.~\eqref{eq:model:full3ode:maintext} borrowing the methodology from statistical physics and chemical kinetics~\citep{vankampen}, which relies on two main assumptions: the system is well mixed and sufficiently large. 
In a well-mixed system, interactions between any pair of units are equally probable independent of space (cf. law of mass action). 
The dynamics of systems that have an interaction network far from a well-mixed state could be accurately modeled through more complex models that explicitly include information about the interaction network structure, \textit{e.g.}, by addressing node degrees and neighborhood sizes~\citep{shang14,khaluf17}. 
The second assumption is that the system is sufficiently large to ignore the effect of finite-size fluctuations. 
This assumption could also be relaxed through more complex models, for instance by formulating a master equation---a macroscopic model that explicitly takes into account stochastic fluctuations with magnitudes proportional to system size~\citep{Gillespie2013}---that has already been successfully employed to model multi-core systems~\citep{Bogdan2011}.
The master equation approach was also applied, for example, to multi-agent systems~\citep{lerman2001general} and traffic flow~\citep{mahnke1999stochastic}.

\subsection{Estimating the performance} 
\label{sec:performace}
The ODE system of
Eq.~\eqref{eq:model:full3ode:maintext} and its fixed points, Eq.~\eqref{eq:fixedPoints}, indicate how the units of the system divide
into the different work states~$S$, $G$, and~$F$. Depending on their
work state, the units contribute differently to the task execution. In
our analyses, we assume that units contribute linearly with a tuneable
coefficient because nonlinear effects may be incorporated in Eq.~\eqref{eq:model:full3ode:maintext}. However, other contribution functions can be tailored to
the performance returns of the specific application scenario.
To model the system's performance in terms of throughput~$X$, we define a function that sums the
contribution of the units in each state as
\begin{equation}
  X(N) = X(s^*,g^*,f^*)= X(s^*,g^*) = c_s s^*+c_g g^*\, , 
  \label{eq:performance-function:throughput}
\end{equation}
where the contribution coefficients $c_s$ and $c_g$ indicate the
contributions to the task execution by units in the two states~$S$ and~$G$, and $(s^*,g^*,f^*)$ is the stable fixed point. 
We assume that units in state \textit{fermo}~$F$ do not contribute to task completion because they are in congestion or lack coherency. 
The system's speedup can be computed as \begin{align}
    S_\text{thruput}(N) &= \frac{X(N)}{X(1)} = \frac{X(s^*,g^*,f^*)}{X(1,0,0)} =
    \frac{c_ss^*+c_gg^*}{c_s}\notag\\  &=
    s^*+\frac{c_g}{c_s} g^*\, ,
  \label{eq:performance-function:speedup}
\end{align}
for $c_s\ne 0$. The game changer is whether $\frac{c_g}{c_s}$ is greater or smaller than one. If $\frac{c_g}{c_s}>1$, collaboration is advantageous and one prefers to have units in state~$G$. Otherwise one would prefer units in state~$S$.
The contribution
coefficients can also be zero as, for example, depending on the task, units in 
state \textit{grupo} may not contribute to task completion. 
For $c_g=0$ we have the standard parallelization problem where we want to keep interactions between units minimal because units in state~$G$ do not contribute to task completion.
The special case~$c_s=0$ can be understood by studying $\lim_{c_s\rightarrow 0}S_\text{thruput}(N)$ that results in an infinite improvement of throughput: $S_\text{thruput}(N)\rightarrow \infty$ for $c_g>0,\,g^*>0$, and~$c_s \rightarrow 0$. Imagine a task that cannot be solved by one robot (\textit{solo} unit) but by two collaborating robots (\textit{grupo} units). We would get $X(1)=0$ and $X(2)=1$, hence, an infinite speedup.
Fig.~\ref{fig:stateMachineAndPerformance} shows a graphical representation
of how the proportion of units in each state is used to compute the
system throughput~$X(s^*,g^*,f^*)$. 

To model systems that benefit from naive parallelization, it is
sufficient to set the contribution coefficients to~$c_s>0$
and~$c_g=0$. In such a contribution scheme, only \textit{solo} workers
contribute to task completion. 
This scheme applies to parallel
computing in which, typically, the ideal case corresponds to having
all the processors operating independently with minimal interaction and hence minimal communication costs~\citep{grama2003introduction,ballard2011minimizing}. 
Consequently, the best performance is attained by
limiting interaction and keeping as many processors as possible in the
state \textit{solo}~$S$.
However, interactions cannot always be avoided and, in certain cases,
especially in robotics, exploiting interaction between units can be
advantageous, or even essential, to the task completion. Examples of
tasks, studied in the literature, that require collaboration in order
to be completed are collective transport~\citep{berman11b} and
collective manipulation~\citep{ijspeert01b} of bulky objects, and
self-assembly of robots to cross difficult
terrain~\citep{ogrady07}. The performance function for tasks that
benefit from interaction has the contribution
coefficient~$c_g>0$. Therefore, the ratio~$c_g/c_s$ determines which
state is more productive.

\subsection{System size and density}

The system size $N$ represents the amount of units (\textit{e.g.},
processors, robots, users) comprising the system.
In multi-robot systems, the robots operate in a physical space in
which physical interference (\textit{e.g.} collisions) typically plays
a determining role in the task execution and may have an impact on the
performance. As one could expect, physical interference is determined
by both the number of robots~$N$ and the size of the working space~$A$. Therefore, to study the scalability of multi-robot systems, it is
appropriate to also consider the impact of the system density~$\rho=N/A$, in addition of the system size~\citep{hamann2018book}. In
this study, we assume a constant working space area~$A=1$ which allows
us to appreciate the effect of both density and system size by only
varying~$N$.


\section{Scenarios}
\label{sec:results}

In scalability analysis, three regimes have been identified and
described through Gunther's Universal Scalability Law~\citep{gunther93}:
\begin{enumerate}
\item ideal concurrency, also described by Gustafson's law
  (Fig.~\ref{fig:gustafson})
\item contention-limited, that corresponds to Amdahl's law
  (Fig.~\ref{fig:amdahl})
\item diminishing returns, for example as seen in Fig.~\ref{fig:gunther}.
\end{enumerate}
In the following we discuss these three scalability regimes and show
that, through specific parameterizations, our general model can
describe 
all of them. Additionally, our model has the added benefit
of establishing a causal link between the scalability regime of the
system and the microscopic behavior of its units. In the following
representative scenarios, we can obtain the three different regimes by
varying the transition rates $k_i$, while keeping constant
the contribution coefficients to $c_s=1$ and $c_g=0$, \textit{i.e.}
only \textit{solo} units (in state $S$) contribute to the execution of
the task. Similar results can be obtained with different contribution
coefficients. Additionally, in Sec.~\ref{sec:superlinear} we show that
superlinear increase can be described by parameterizing contribution
coefficients to values larger than one. Finally, without loss of
generality, we assume that the system is initialized at time~$t=0$
with only \textit{solo} units, \textit{i.e.} $s(0)=N, g(0)=0, f(0)=0$.



\subsection{Ideal concurrency -- Gustafson's law}
\label{sec:ideal}

Ideal concurrency, as described by Gustafson's law
(Eq.~\eqref{eq:gustafson}), corresponds to the scenario in which
increasing the system size leads to unbounded increase in
performance. The optimal case consists in a linear increase of the
performance as $S_\text{thruput}=N$. A~situation where units
do not interact and hence act independently from each
other. In Gustafson's law of Eq.~\eqref{eq:gustafson}, this situation
is obtained by setting the serial proportion parameter~$\sigma=0$, and
similarly, in Gunther's USL of Eq.~\eqref{eq:usl} by setting both the
contention and coherency delay parameters $\sigma=\kappa=0$.
In our model, the lack of any interaction can be parameterized by
setting all the transition rates to zero ($k_i=0$ for
$i\in\{1,2,\dots,7\}$). In this case, the stable point is
$(s^*=N, g^*=0, f^*=0)$ 
and thus
$P(s^*, i^*, j^*)=N$. 

\begin{figure}[t]
    \centering
    \begin{subfigure}[c]{0.47\textwidth}
      \includegraphics[width=1\textwidth]{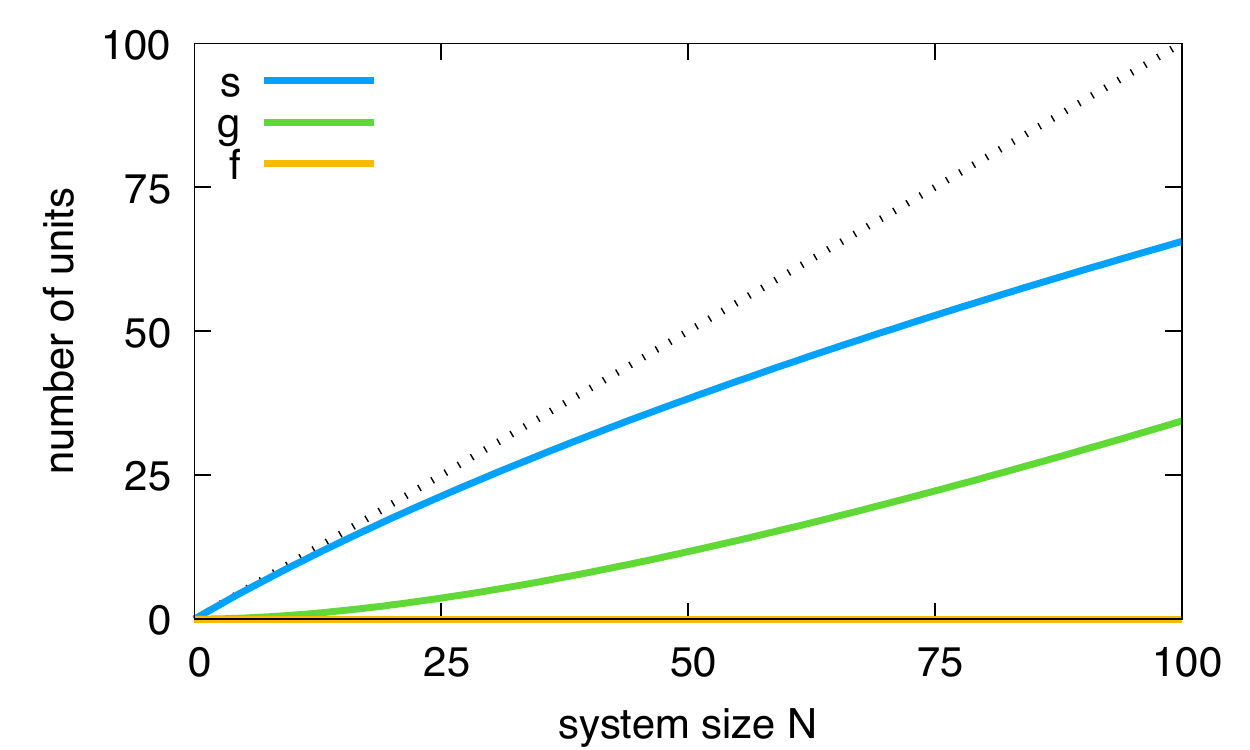}
      \subcaption{\label{fig:ideal-concurrency:pop}ideal concurrency (Gustafson's law), states of units ($k_1=0.004$, $k_4=1.0$)}
    \end{subfigure}
        \hspace{2mm}
    \begin{subfigure}[c]{0.47\textwidth}
      \includegraphics[width=1\textwidth]{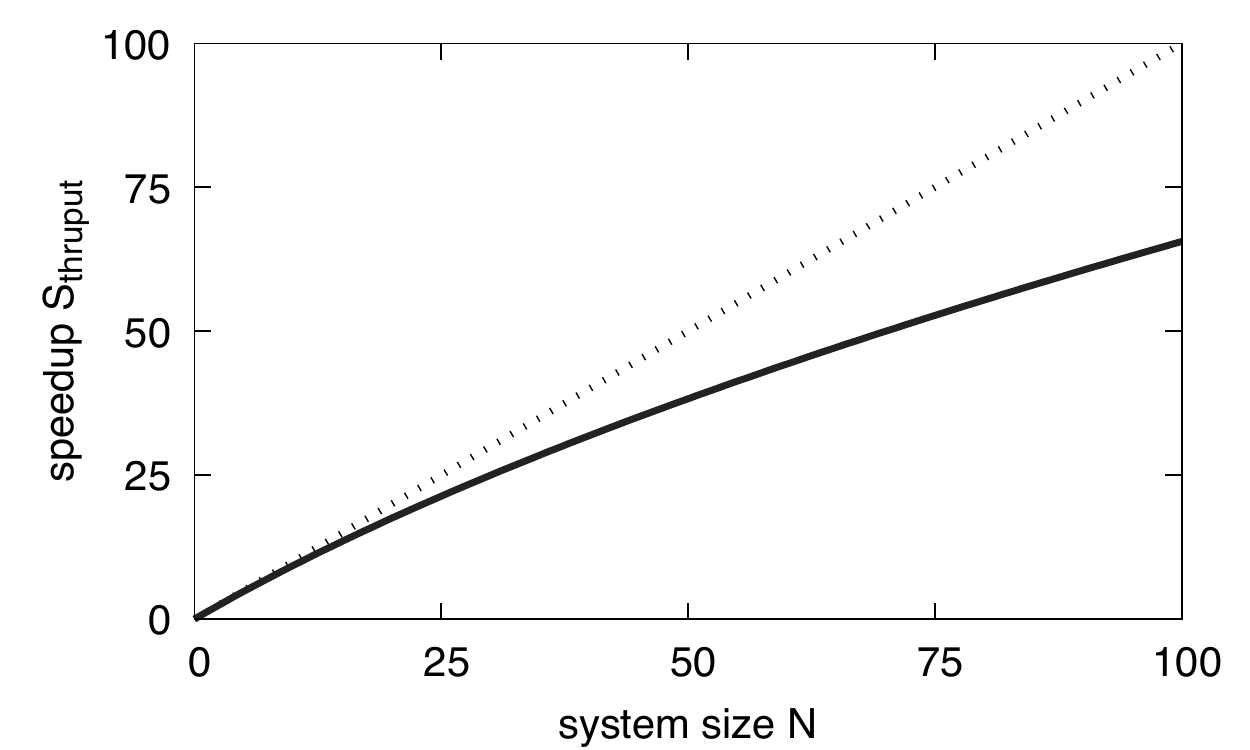}
      \subcaption{\label{fig:ideal-concurrency:p}ideal concurrency (Gustafson's law), speedup ($k_1=0.004$, $k_4=1.0$, $c_s=1$, $c_g=0$)}
    \end{subfigure}    
    \begin{subfigure}[c]{0.47\textwidth}
      \includegraphics[width=1\textwidth]{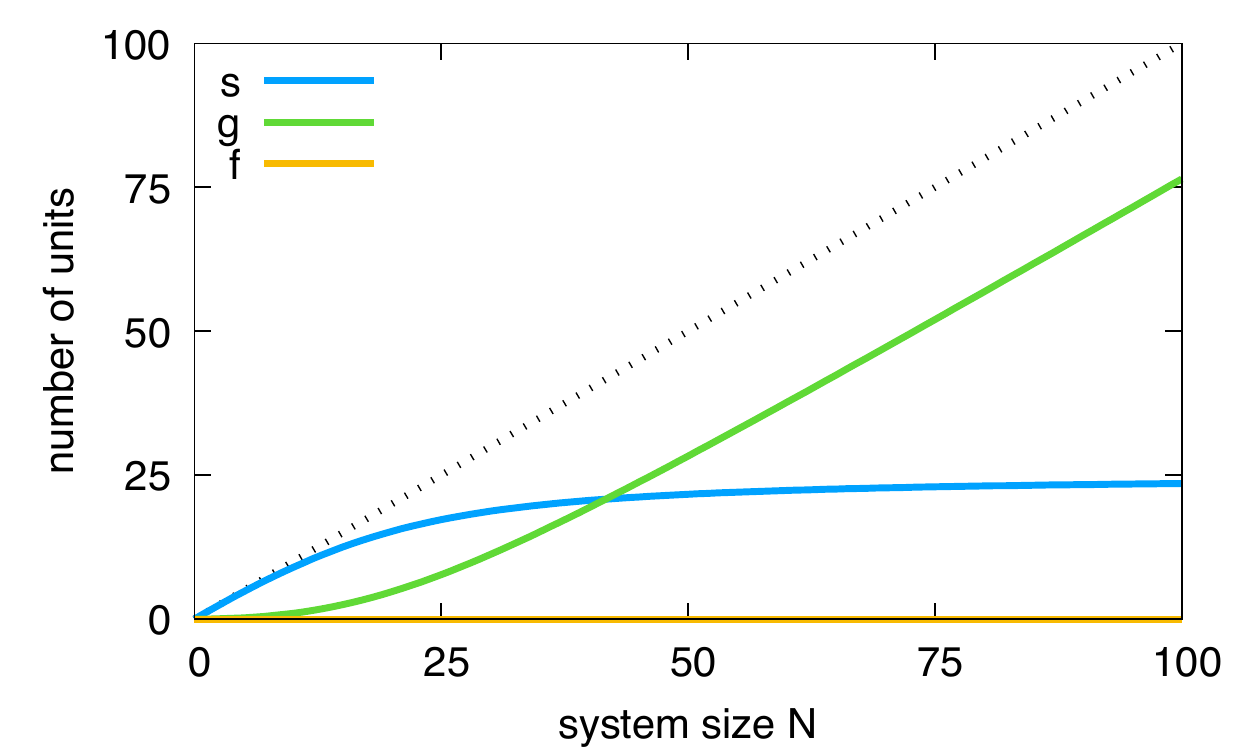}
      \subcaption{\label{fig:contention-limited:pop} contention-limited
        (Amdahl's law), states of units ($k_1=0.004$, $k_2=0.04$, $k_4=1.0$)}
    \end{subfigure}
    \hspace{2mm}
    \begin{subfigure}[c]{0.47\textwidth}
      \includegraphics[width=1\textwidth]{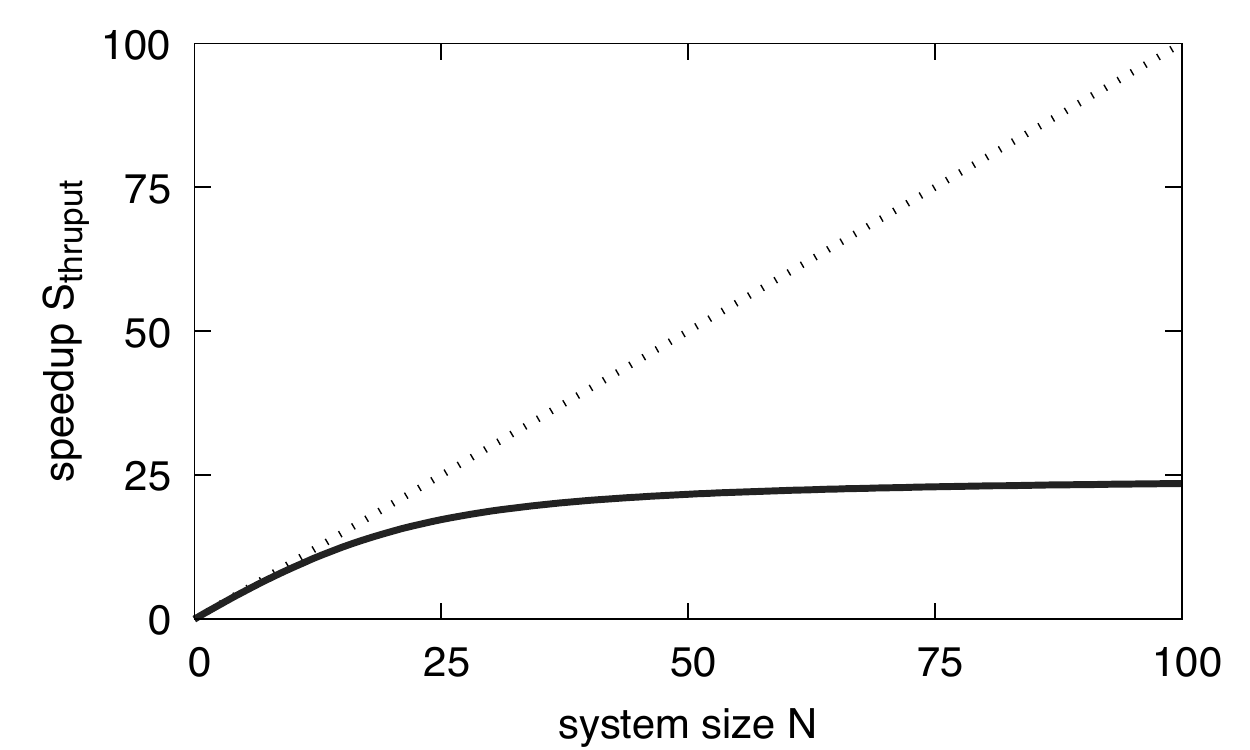}
      \subcaption{\label{fig:contention-limited:p} contention-limited
        (Amdahl's law), speedup ($k_1=0.004$, $k_2=0.04$, $k_4=1.0$, $c_s=1$, $c_g=0$), $\sigma=k_2/k_4=0.04$}
    \end{subfigure}    
    \caption{Our model is able to reproduce scalability dynamics
      described by the existing models:
      (a, b)~Gustafson's law, and
      (c, d)~Amdahl's law. Plots on the left
      show the number of units by their states
      with the increase of the system size~$N$. Plots on the right show the
      speedup~$S_\text{thruput}$ as a function of~$N$. In the ideal
      concurrency situation, (a, b), $S_\text{thruput}$ unboundedly increases by increasing~$N$; in the contention-limited regime, (c, d), $S_\text{thruput}$ saturates to a
      constant value for large~$N$. The value of the rates are
      indicated below the plots, unspecified rates are set to
      zero.}
    \label{fig:regime1-2}
\end{figure}

In Gustafson's law of Eq.~\eqref{eq:gustafson}, when the serial
proportion parameter $\sigma>0$, the system performance keeps
increasing with $N$, however at sublinear speed. Our model can describe
this scenario by setting the rates $k_1 >0$ and $k_4>0$, while keeping
$k_i=0$ for $i\in\{2,3,5,6,7\}$. In this case, by computing the stable
fixed point of Eq.~\eqref{eq:model:full3ode:maintext} (see Appendix)
we obtain
\begin{equation}
  S_\text{thruput}(N) =\frac{\sqrt{k_4^2 + 8 k_1 k_4 N}-k_4}{4 k_1}\,,
  \label{eq:fp:gustafson}
\end{equation}
which approximates the dynamics of Gustafson's law for $k_4>k_1$.
Figure~\ref{fig:ideal-concurrency:p} shows a representative configuration
of Eq.~\eqref{eq:fp:gustafson} for $k_1=0.002$ and $k_4=1$. Similarly
to Gustafson's law, Eq.~\eqref{eq:fp:gustafson} sublinearly grows with
the system size $N$ without saturating, even for large $N$.

\subsection{Contention-limited -- Amdahl's law}
\label{sec:contentionLimited}

Amdahl's law (Eq.~\eqref{eq:amdahl}) describes the scalability
dynamics where the performance increase is limited by interaction
cost. In fact, $S_\text{thruput}$ increases with the system size $N$ until a
saturation limit, after which the performance stays constant with
increasing system size $N$. Gunther's USL (Eq.~\eqref{eq:usl}) can
also describe such dynamics by setting the contention parameter
$\sigma>0$ and the coherency delay parameter $\kappa=0$. Our model
reproduces such dynamics for $k_1>0$, $k_2 >0$ and $k_4 >0$, while
keeping $k_i=0$ for~$i\in\{3,5,6,7\}$ (see details in the
Appendix). As a difference from the ideal concurrency case
(Sec.~\ref{sec:ideal}), in this scenario the interference has a
relatively strong positive feedback (rate $k_2$) that limits the
increase of performance with $N$.

In the Appendix, we show that for the special case~$k_2=2k_1$ the stable fixed point $(s^*, g^*, f^*)$
 yields
\begin{equation}
  S_\text{thruput}(N)=\frac{N}{1+\frac{k_2}{k_4}N} \;\,.
  \label{eq:fp-amdhal-special}
\end{equation}
This equation has the same form as Amdahl's law of
Eq.~\eqref{eq:amdahl} suggesting that the serial part parameter
$\sigma=\frac{k_2}{k_4}$. On the one hand, increasing rate~$k_4$
in our model corresponds to decreasing the serial part of the task in
Amdahl's law. In our model, $k_4$ dictates how frequently units
complete interactions (\textit{i.e.}, leave state \textit{grupo}~$G$ and return to
\textit{solo}~$S$), and by increasing~$k_4$, the interaction cost decreases. 
On the
other hand, the interaction feedback~$k_2$ increases the rate of units
moving to state~$G$, and by removing it ($k_2=k_1=0$) the model
reduces to the ideal concurrency case of Sec.~\ref{sec:ideal}.

Figs.~\ref{fig:contention-limited:pop} and~\subref{fig:contention-limited:p} show the results for $k_1=0.004$,
$k_2=0.04$, and $k_4=1.0$ that resemble Amdahl's law dynamics of
Fig.~\ref{fig:amdahl}. Note that in region $N < \frac{k_4}{k_2}$ the performance does not saturate because units complete interactions at a rate faster
than interactions take place, leading to a permanently increasing performance. For $N > \frac{k_4}{k_2}$ the curve $S_\text{thruput}(N)$ starts flattening, and for $N\rightarrow \infty$ Eq.~\ref{eq:fp-amdhal-special} converges to $S_\text{thruput}(N)\approx \frac{k_4}{k_2}$.

\subsection{Diminishing returns}
\label{sec:diminishing}

In the diminishing returns regime, the system increases in performance
with increasing~$N$ until the critical value~$N_c$. Increasing the
system size beyond the critical value, $N>N_c$, leads to a decrease in
performance due to an increase of the coordination costs.
Gunther's USL describes such dynamics with the coherency delay
parameter $\kappa>0$ (regardless of the contention parameter's value
$\sigma$). In our model, the diminishing returns regime is the result
of feedbacks by units in states \textit{grupo}~$G$ and \textit{fermo}~$F$ that cause the transition
of more and more units to these (here unproductive) states. By setting
every rate $k_i>0$ for $i\in\{1,...,7\}$, the system can go in a
diminishing returns regime. While we were able to find the symbolic
function of the stable fixed points, they were intractable because they are too long
and cannot provide any analytical benefit. However, in the Appendix,
we show that fixing the rates $k_2 =k_3 =k_5 =k_6 =2k_1$ and
$k_4 =k_7$, and through an approximation, we find 
\begin{equation}
  S_\text{thruput}(N) = \frac{N}{\frac{k_2^2}{k_4^2}N^2+\frac{k_2}{k_4}N+1} \;.
  \label{eq:approxUSL}
\end{equation}
Interestingly, our model as given in Eq.~\eqref{eq:approxUSL} and being derived from a
microscopic model of interactions among units is similar
to Gunther's USL of Eq.~\eqref{eq:usl}. The parameters are
linked as $\sigma=\frac{k_2}{k_4}$ (in line with what has been
discussed in Sec.~\ref{sec:contentionLimited}) and
$\kappa = \frac{k_2^2}{k_4^2} = \sigma^2$. The parameters~$\kappa$ and~$\sigma$ are not restricted in this way in the original USL. Gunther et al.~\citep{gunther15} allow
both positive and negative contention parameters~$\sigma$,
depending on the system's characteristics, to describe positive or
negative contention (see also Sec.~\ref{sec:superlinear}). 
By argument of symmetry one can ask if also the coherency delay parameter~$\kappa$ should
accept negative values to represent a form of `negative lack of
coherence', or in other words, a form of super-coherence. 
However, according
to Eq.~\eqref{eq:approxUSL} we have
$\kappa = \frac{k_2^2}{k_4^2} = \sigma^2$ prohibiting negative
values.
Fig.~\ref{fig:diminishingReturns} shows an example of the diminishing returns
regime. Such system dynamics have been reported in parallel computing
\citep{gunther93} as well as in multi-robot
scenarios~\citep{hamann18b}.

\subsection{Superlinear speedup}
\label{sec:superlinear}

\begin{figure}[ht]
    \centering
    \begin{subfigure}[c]{0.47\textwidth}
      \includegraphics[width=1\textwidth]{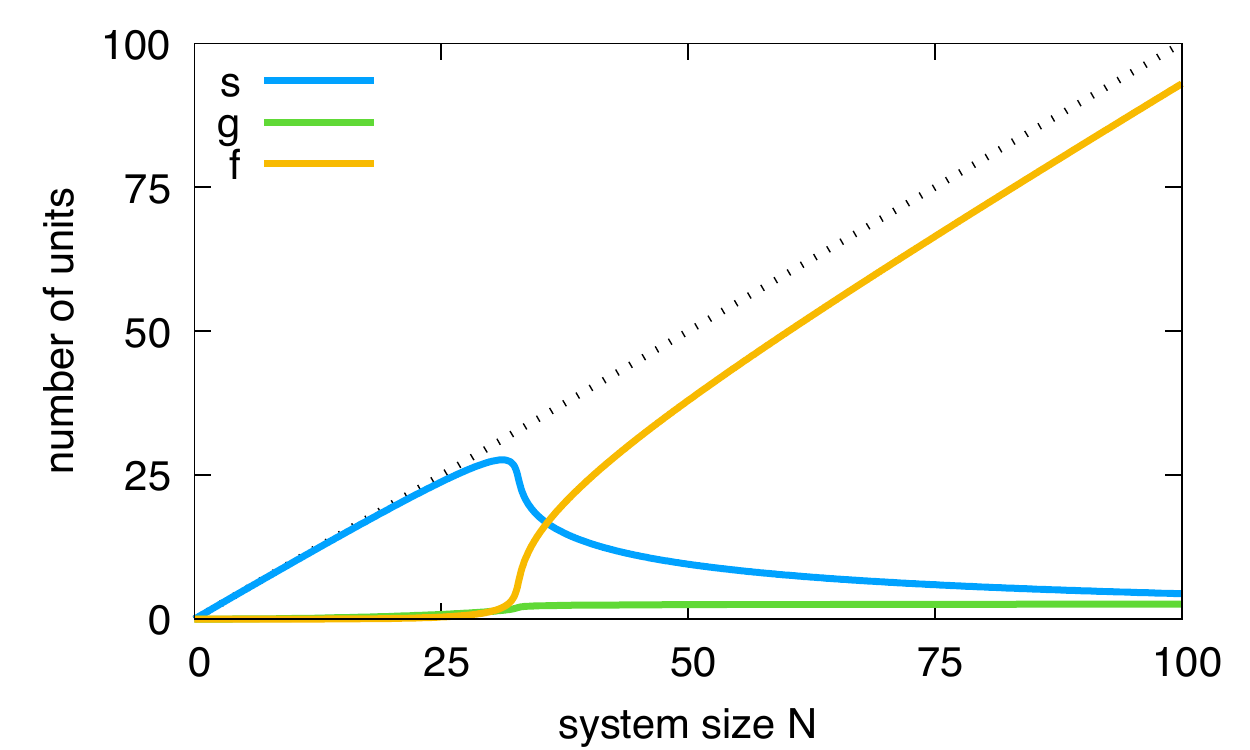}
      \subcaption{\label{fig:diminishingReturns:pop}states of the units}
    \end{subfigure}
     \begin{subfigure}[c]{0.49\textwidth}
      \includegraphics[width=1\textwidth]{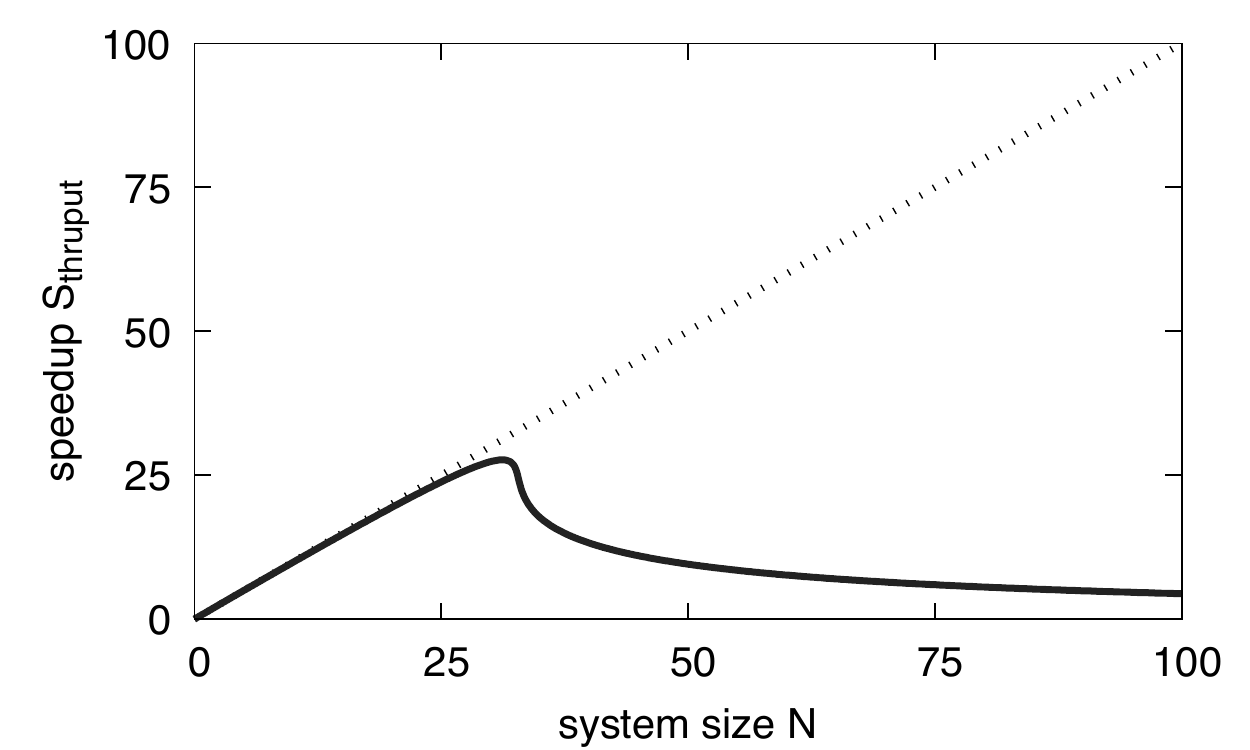}
      \subcaption{\label{fig:diminishingReturns:p1}speedup $S_\text{thruput}$ based on \textit{solo}~$S$, $c_s=1$, $c_g=0$}
    \end{subfigure}
    \hspace{2mm}
    \begin{subfigure}[c]{0.47\textwidth}
      \includegraphics[width=1\textwidth]{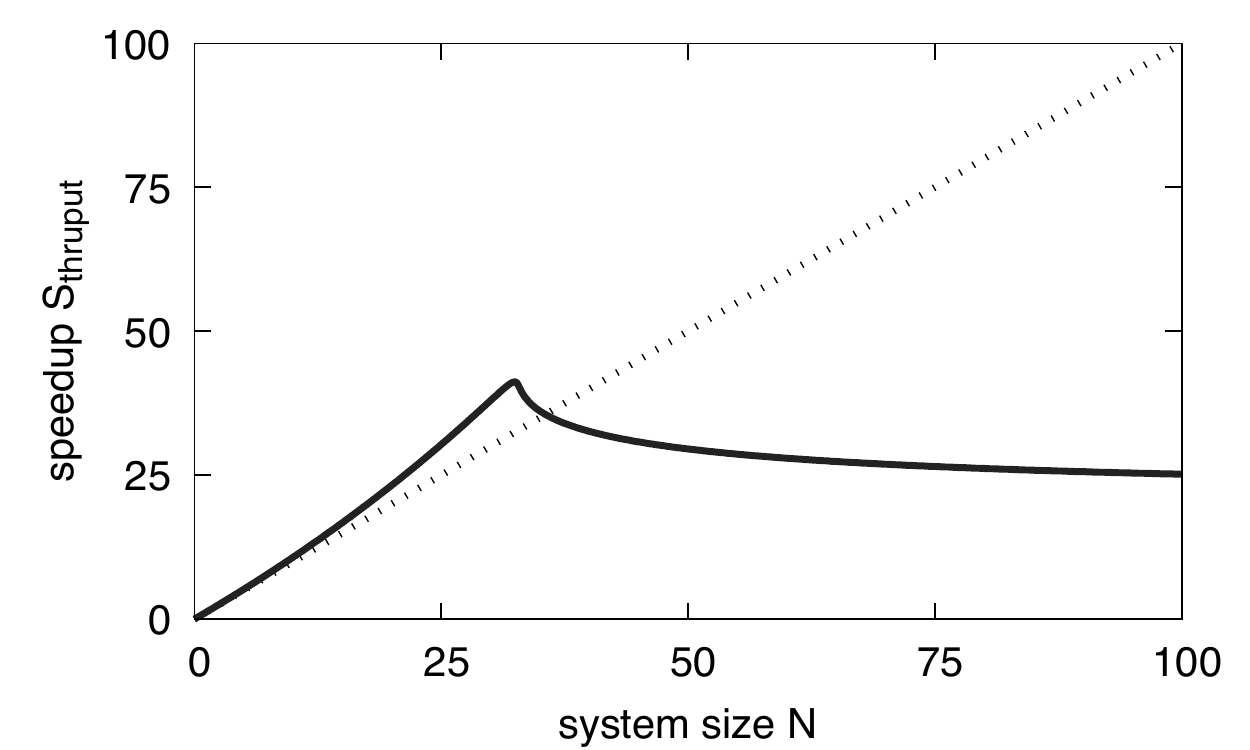}
      \subcaption{\label{fig:diminishingReturns:p2}speedup $S_\text{thruput}$ based on \textit{solo}~$S$ and \textit{grupo}~$G$, $c_s=1$, $c_g=8$}
    \end{subfigure}
    \caption{The same population dynamics (panel (\subref{fig:diminishingReturns:pop}) with rates from Tab.~\ref{tab:diminishingReturnsParameters}) can give (\subref{fig:diminishingReturns:p1}) diminishing returns or (\subref{fig:diminishingReturns:p2}) superlinear speedup, depending on the coefficient parameters. In (\subref{fig:diminishingReturns:p1}), $c_s=1$, $c_g=0$ indicate that the system benefits from \textit{solo} units only. Instead, in (\subref{fig:diminishingReturns:p2}), $c_s=1$, $c_g=8$ indicate that units in state \textit{grupo} are eight times more efficient than \textit{solo} units, leading to a superlinear speedup (when the curve is above the dashed diagonal).}
    \label{fig:diminishingReturns}
\end{figure}

As reported by Gunther et al.~\citep{gunther15}, certain systems can benefit from 
interactions between units to achieve a superlinear speedup of
performance. Such a scenario is particularly relevant in multi-robot
systems in which robots may be able to cooperate with each other to
improve their performance to a greater extent than they would do by
executing the task independently (\textit{e.g.}, \citep{berman11b,
  ijspeert01b, ogrady07, Talamali:SwInt:2019}). For example, in a
collective resource collection task (\textit{e.g.}, based on pheromone
trails \citep{Talamali:SwInt:2019}), a single robot may take long time
to find the unknown location of the resource cluster. Conversely,
groups of robots can share information on individual
searches and superlinearly speedup the collective task.

In our model, the beneficial or detrimental nature of the interactions
is specified via the contribution coefficients. 
Values of~$c_g$
that are larger or smaller than~$c_s$ indicate that units are more productive
in state \textit{grupo} or \textit{solo}, respectively. 
With~$c_g<c_s$, each
unit in state~$G$ is less effective than units in state~$S$, therefore
interactions should be minimized to improve the performance. On the
contrary, $c_g>c_s$ indicates that interacting units, in state~$G$,
are more effective than units operating independently, therefore
interactions should be maximized to improve performance. Note that
how beneficial (or detrimental) states are, is independent from
the details of the population dynamics (transition rates~$k_i$). 
Fig.~\ref{fig:diminishingReturns} shows such an example: the same
underlying population dynamics (Fig.~\ref{fig:diminishingReturns:pop} with parameters of Tab.~\ref{tab:diminishingReturnsParameters}) can lead to two different speedup curves (Figs.~\ref{fig:diminishingReturns:p1} and~\ref{fig:diminishingReturns:p2}) as a consequence of different contribution coefficients.

\begin{table}[t]
\centering
\begin{tabular}{ll}
$k_1$     &  0.005  \\
$k_2$     &  0.1  \\
$k_3$     &  0.06  \\
$k_4$     &  10 \\
$k_5$     &  0.15  \\
$k_6$     &  0.3 \\
$k_7$     &  0.8  
\end{tabular}
\caption{\label{tab:diminishingReturnsParameters}Transition rates for diminishing returns and superlinear speedup shown in Fig.~\ref{fig:diminishingReturns}.}
\end{table}

Superlinear speedups are observed when beneficial interactions among units cause the performance of a system of
$N$ units to be $S_\text{thruput}>N$. Gunther's USL can model superlinear
speedups by setting the contention parameter to negative values,
$\sigma<0$, as it represents beneficial interactions. Our model can
show superlinear speedups when the ratio of the contribution coefficients $\frac{c_g}{c_s}$ is larger than
one, as indicated in Eq.~\eqref{eq:performance-function:speedup}. When the \textit{grupo} coefficient~$c_g$ is larger than the \textit{solo} coefficient~$c_s$, then collaborating units in state \textit{grupo} are more effective than \textit{solo} units. Therefore,
the system throughput of~$N$ units can be greater than $N$-times the throughput $c_s$ of a single unit. In our model, superlinear speedup is a combination of the task's characteristic---that is, the benefits of collaboration---and of the system's functioning---that is, the number of units that collaborate with each other.

\section{Showcase: model fitting}
\label{sec:fitting}

\begin{figure}[h!]
    \centering
    \begin{subfigure}[c]{0.48\textwidth}
      \includegraphics[width=1\textwidth]{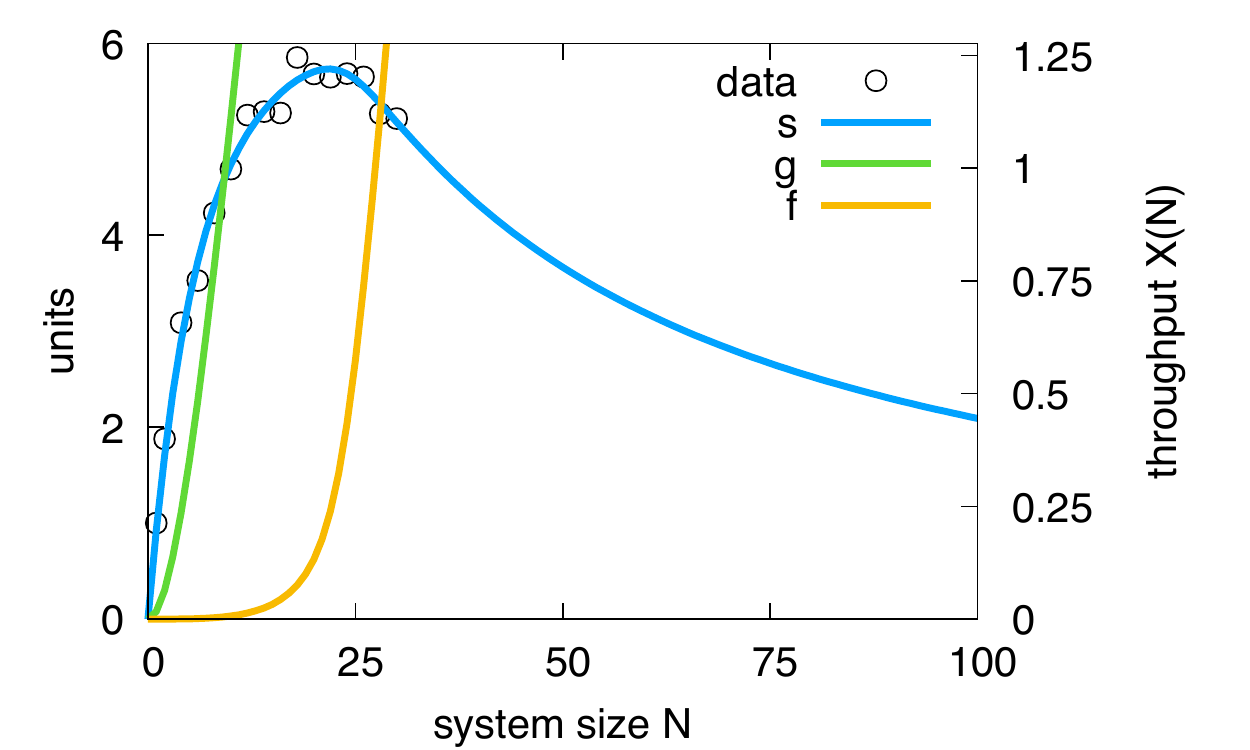}
      \subcaption{\label{fig:fitting:a}SQL Server, dataset from Gunther~\citep{gunther07}}
    \end{subfigure}   
    \hspace{2mm}
    \begin{subfigure}[c]{0.48\textwidth}
      \includegraphics[width=1\textwidth]{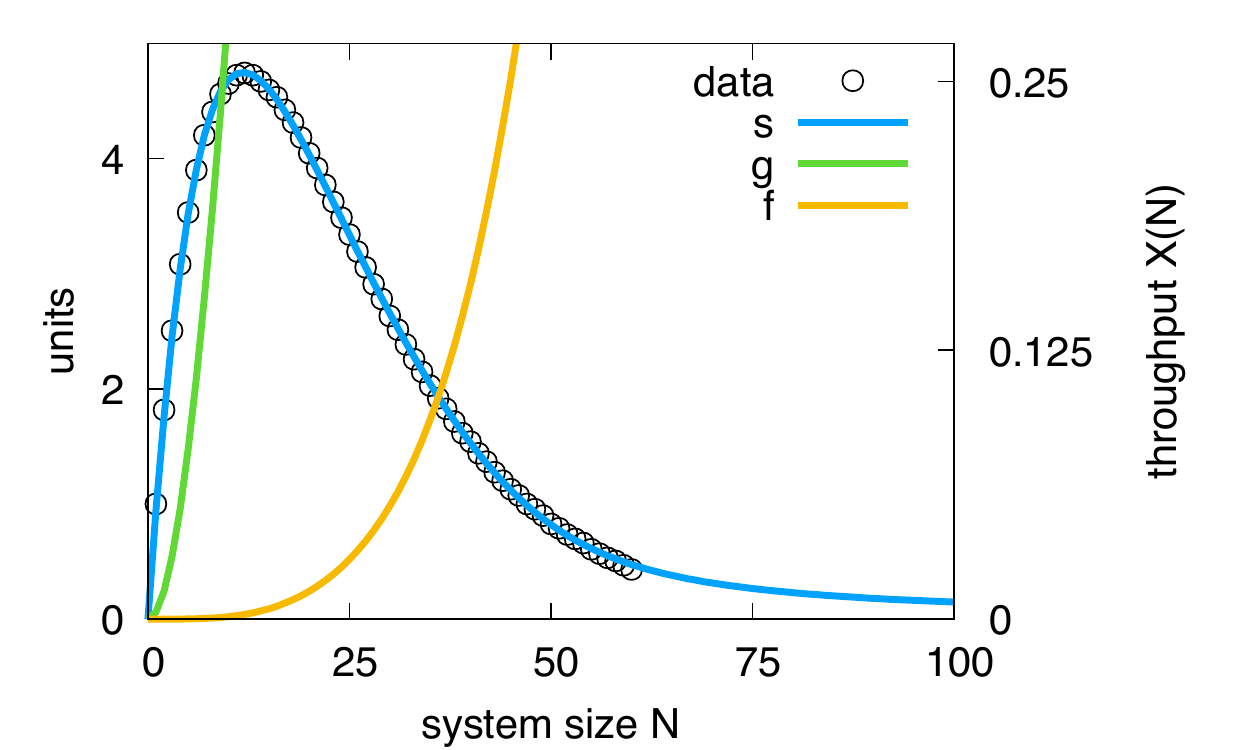}
      \subcaption{\label{fig:fitting:b}wireless network with ALOHA protocol, dataset  from Gokturk et al.~\citep{gokturk08}}
    \end{subfigure}
    \begin{subfigure}[c]{0.48\textwidth}
      \includegraphics[width=1\textwidth]{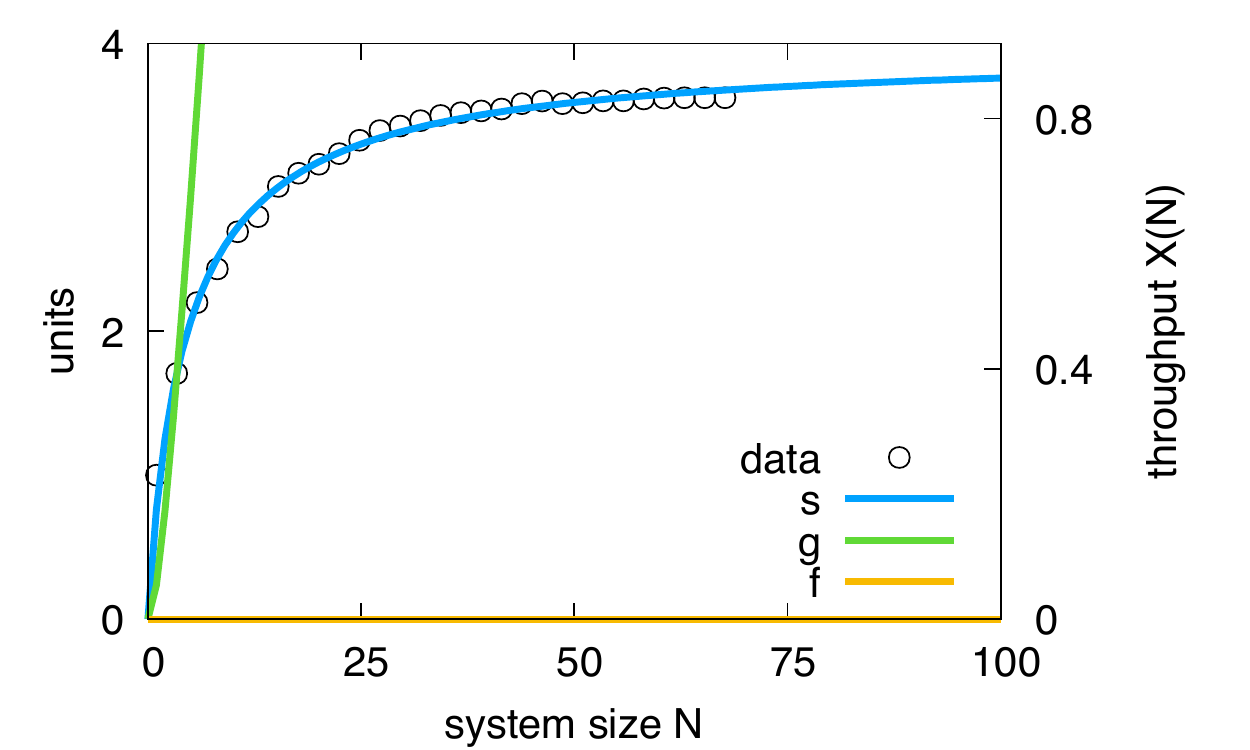}
      \subcaption{\label{fig:fitting:c}foraging self-propelled particles (going through each other), dataset from Rosenfeld et al.~\citep{rosenfeld06}}
    \end{subfigure}    
    \hspace{2mm}
    \begin{subfigure}[c]{0.48\textwidth}
      \includegraphics[width=1\textwidth]{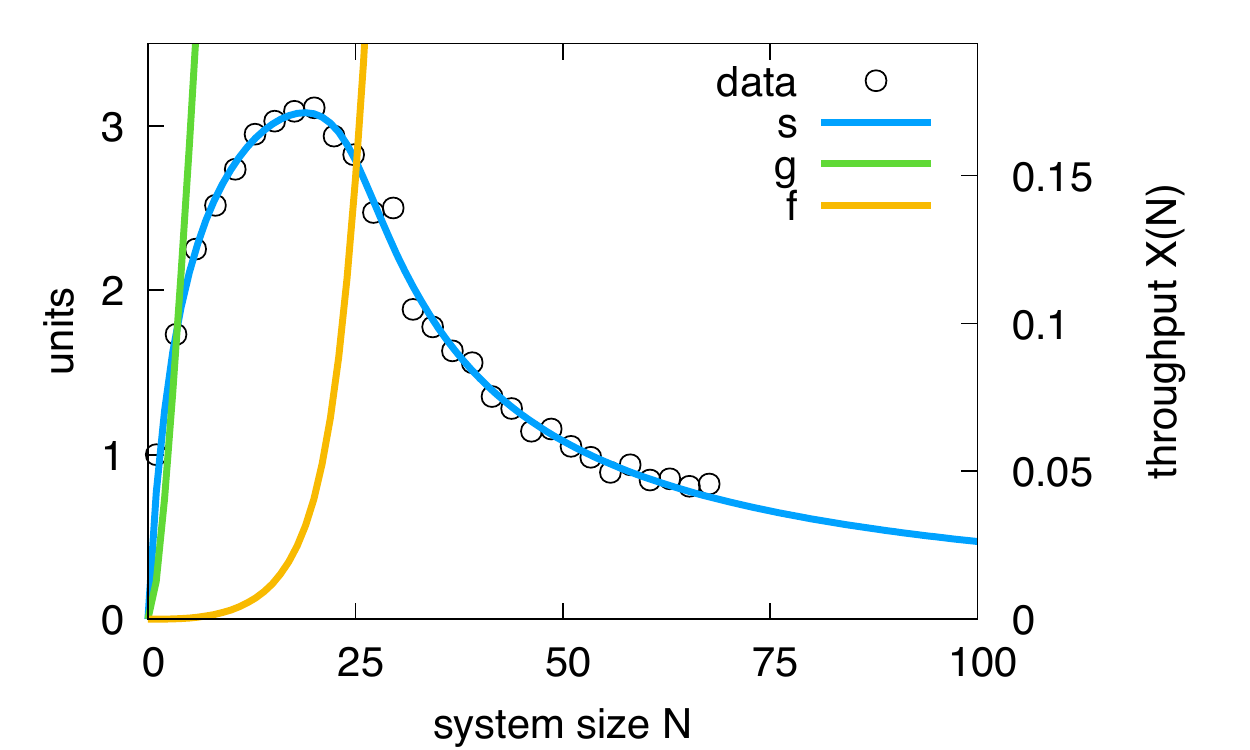}
      \subcaption{\label{fig:fitting:d}foraging robot swarm (embodied system), dataset from Rosenfeld et al.~\citep{rosenfeld06}}
    \end{subfigure}
    \begin{subfigure}[c]{0.48\textwidth}
      \includegraphics[width=1\textwidth]{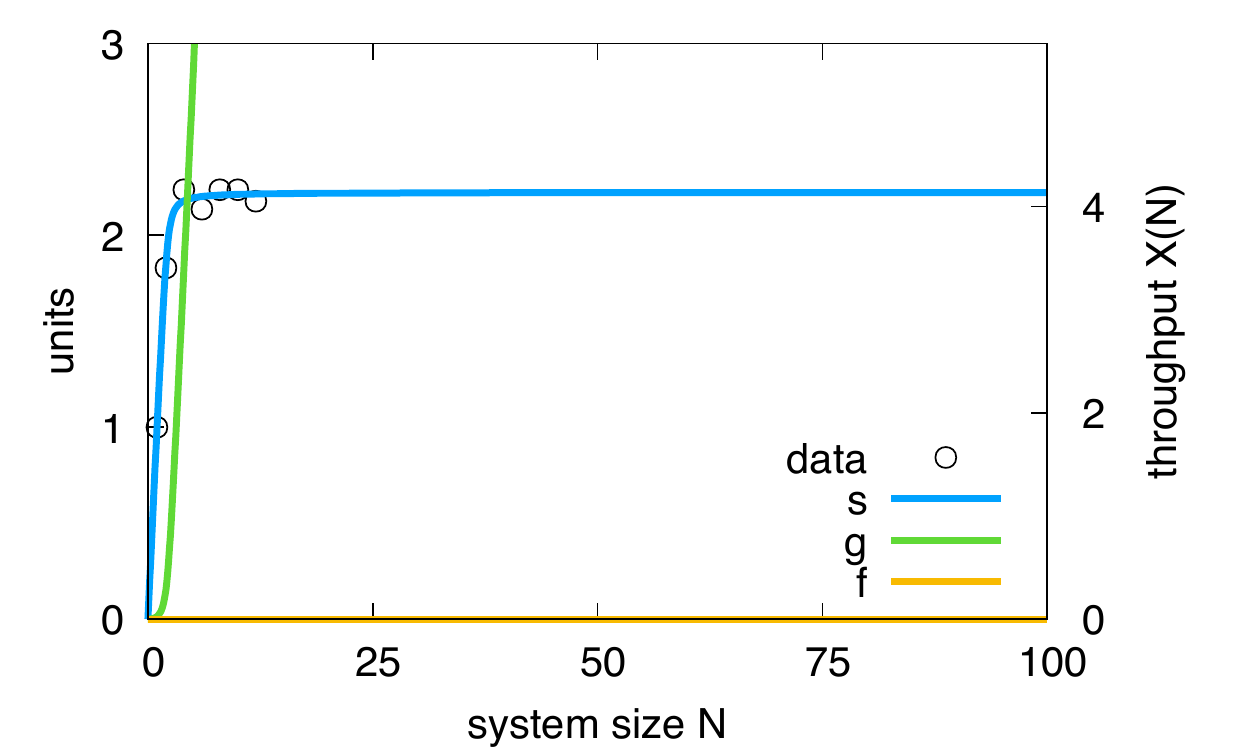}
      \subcaption{\label{fig:fitting:e}NAS parallel benchmark, Block Tridiagonal (BT), dataset from Ribeiro et al.~\citep{ribeiro12}}
    \end{subfigure}
    \hspace{2mm}
    \begin{subfigure}[c]{0.48\textwidth}
      \includegraphics[width=1\textwidth]{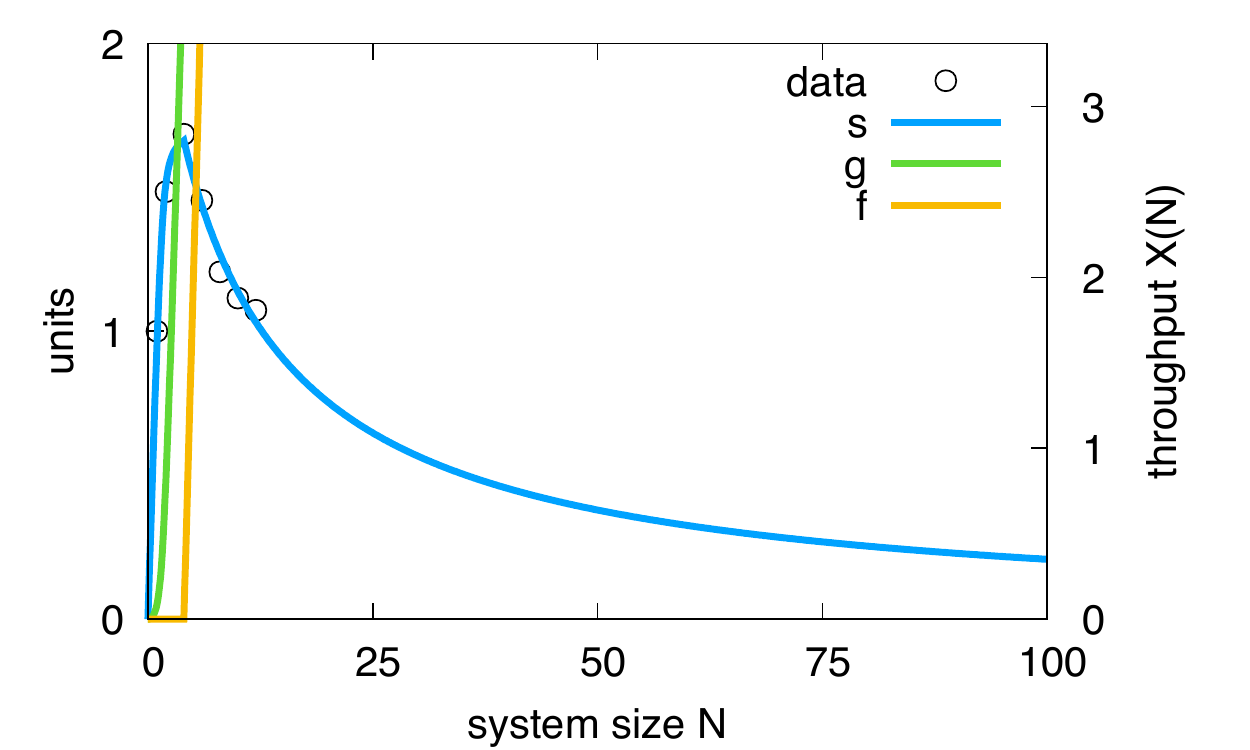}
      \subcaption{\label{fig:fitting:f}NAS parallel benchmark, Scalar Pentadiagonal (SP), dataset from Ribeiro et al.~\citep{ribeiro12}}
    \end{subfigure}
    \caption{Fitting Eqs.~\ref{eq:model:full3ode:maintext} and~\ref{eq:performance-function:throughput} to datasets from
    parallel computing~\citep{gunther07},
    wireless networks~\citep{gokturk08}, and swarm
    robotics~\citep{rosenfeld06}. For simplicity all horizontal axes are normalized to the range $N \in [1,100]$. The fitted parameters are given in Tab.~\ref{tab:fitting}.}
    \label{fig:fitting}
\end{figure}

In order to showcase the wide applicability of the proposed approach,
we fit our scalability model to published data from three different
fields: parallel computing, wireless sensor networks, and multi-robot
systems. Obtaining a good fit is an indicator of how the model could
potentially reproduce the dynamics observed in parallel and distributed
systems. In the considered systems, the task is performed by 
units in state \textit{solo} only, therefore we set the contribution
coefficients of Eq.~\ref{eq:performance-function:throughput} to
$c_s>0$ and~$c_g=0$. The value of $c_s$ is determined for each task as $X(1)=c_s$ (following the discussion in Sec.~\ref{sec:performace}).
While we set the contribution coefficients this way by
hand based on our interpretation of the tasks, we automated the
identification of the transition rates of
Eqs.~\eqref{eq:reaction1} to~\eqref{eq:reaction7} through a fitting
routine. 
We fit the performance function~$X(N)$ of
Eq.~\eqref{eq:performance-function:throughput} to the performance data
taken from the literature. As~$X(N)$ depends on the stable fixed point
of the system of Eq.~\eqref{eq:model:full3ode:maintext}, we
numerically computed it through time integration with starting point
$s(0)=N$, $g(0)=0$, $f(0)=0$. As fitting routine, we employed the
differential evolution method~\citep{storn1997differential} implemented in SciPy~\citep{SciPy}.

We show that our model can fit data from the following four distinct
systems: an SQL server benchmark on a multiprocessor machine, a~media
access protocol in a wireless sensor network, a~simulation of
self-propelled particles, and a simulation of a robot swarm performing
the foraging task. In this section, we describe in more detail the
four systems and briefly discuss the obtained fits. The fitted
transition rates for the four studied systems are reported in
Tab.~\ref{tab:fitting}, which also includes the contribution coefficient~$c_s$ used to scale throughput~$X(N)$ to the data
range. Fig.~\ref{fig:fitting} shows the data superimposed to state
curves for the fitted transition rates. While the fit has been done on
the throughput~$X(N)$, we plot the fixed point of the three
states because it gives some additional information about the other states
and, in the studied cases, the plot for the \textit{solo} units is directly proportional
to~$X(N)=c_s\,s^*$ (with $c_s>0$ and~$c_g=0$). For simplicity, we also
normalize system size~$N$ (horizontal axis of Fig.~\ref{fig:fitting}) with linear scaling setting the first data point to~$N=1$.

The data of the SQL server benchmark have been taken
from~\citep{gunther07}. The performance data are the transactions per
second, and the system size is the number of virtual users
($N\in\{1,2,\dots,30\}$). The published data includes results for
different hardware configurations, we focus on the `4-way processor
configuration' data.
The fitting results of Fig.~\ref{fig:fitting:a} indicate a good
fit. Interpreting the fitted parameters of Tab.~\ref{tab:fitting}
is nontrivial. However, we notice that the ratio between rate~$k_4$ and rate~$k_2$ predicts an increase of~$X(N)$ for~$N\le7$ (as $\frac{k_4}{k_2}\approx7$) approximately, as shown in Fig.~\ref{fig:fitting:a}. Additionally, rates~$k_3,k_5,k_6>0$ indicate that units may eventually arrive to state~\textit{fermo} anticipating diminishing returns for large~$N$.

Next, we fit data of a wireless sensor network that applies a common
protocol for media access called ALOHA, the data have been taken
from~\citep{gokturk08}. The performance data is the throughput,
measured as number of packets per slot (numerical results), and the
system size is the number of backlogged nodes. The results shown in
Fig.~\ref{fig:fitting:b} indicate a good fit. By looking at the fitted
transition rates in Tab.~\ref{tab:fitting}, we note that rate~$k_3$ is significantly larger than all other rates (except for~$k_7$). This rate triggers
state transitions from state \textit{solo} to state \textit{grupo} 
depending on the number of units in \textit{fermo}. This rate hence
causes the steep decline in performance for~$N>13$ when media access
quickly gets more difficult due to congestion.

The self-propelled particle system is composed of a group of units
that can move in bidimensional space without any collision among
them. The data have been taken from~\citep{rosenfeld06} (algorithm
\textit{goThru}). The self-propelled particles perform the swarm
robotics benchmark of \textit{foraging}, that is, they are tasked to
collect items scattered in the environment and transport them to a
central location. The
particles locally communicate with each other in order to coordinate
the foraging operation. The performance data are the number of
collected items and the system size is the number of particles. As the
particles can go through each other there is no physical interference
among them. In this case, the limiting resource is the number of
available items that leads the system performance to saturate to a
fixed value for large~$N$, in a way similar to Amdahl's law. Due to the lack of interference, we set $k_5=k_6=0$ to turn off
state~\textit{fermo}, see Tab.~\ref{tab:fitting}. The results shown
in Fig.~\ref{fig:fitting:c} indicate a good fit. As expected, the
curve saturates for large~$N$.

We study another system, this one is a robot swarm. The data have been taken
from~\citep{rosenfeld06} (algorithm \textit{timeRand}). The robots
perform the same foraging task of the previous system. However, in
this system, the robots cannot go through each other's bodies and
therefore they may physically interfere with the movements of one
another. Due to physical interference, the expected scalability curve
is diminishing performance for large system size~$N$, as for large
robot densities, room required to avoid collisions is
limited. The results shown in Fig.~\ref{fig:fitting:d} indicate a good
fit. Parameters~$k_5$ and~$k_6$ (transition~$G\rightarrow F$) are small relative to~$k_1, k_2, k_3$ to model
the late but steep decline on the interval~$24<N<60$. Note that the shapes
of~$X(N)$ in Fig.~\ref{fig:fitting:b} and Fig.~\ref{fig:fitting:d}
seem similar but in the swarm robotics system there is first a
saturation for~$N\approx 20$ and then a steep decline indicating a
different underlying process.

Finally, we study two more datasets from parallel computing to include more representative parallel applications. 
Both applications are from the NAS Parallel Benchmark~\citep{jin99}.
The benchmark is based on a number of tasks for computational fluid dynamics.
Here, we focus on two of them: BT and SP. 
Task~BT is based on an algorithm to solve 3-dimensional compressible Navier-Stokes equations that form block-tridiagonal (BT) systems.
Task~SP is based on a similar problem but solves scalar pentadiagonal (SP) bands of linear equations.
We rely on benchmark data from Ribeiro et al.~\citep{ribeiro12}. 
They investigate the impact of different architectures for shared memory but we focus only on their data for uniform memory access (UMA) using a machine with four six-core Intel Xeon~X7460 and Linux~2.6.32. 
We fit their data ``Intel UMA'' for BT and SP of speedup over number of cores (see Figs.~\ref{fig:fitting:e} and~\ref{fig:fitting:f}). 
For task~BT, the system performance saturates as seen previously for the self-propelled particle system in Fig.~\ref{fig:fitting:c}. 
According to Ribeiro et al.~\citep{ribeiro12} this may be caused by memory issues and thread migrations. 
Similar to the case of the self-propelled particle system, we set $k_5=k_6=0$ to turn off
state~\textit{fermo}, see Tab.~\ref{tab:fitting}.
For task~SP, We observe diminishing performance for too large system size~$N$. 
According to Ribeiro et al.~\citep{ribeiro12} this may be caused by memory issues and non-continuous communication. 
For both tasks the results in Figs.~\ref{fig:fitting:e} and~\ref{fig:fitting:f} indicate good and reasonable fits despite the few available data points (seven each).

As already discussed, one of the advantages of our model is the link
between the state of units with the performance curve. By
investigating the casual effect of each rate on the system
performance, we can achieve a better understanding of the system
behavior. However, in this paper we leave out any articulated
discussion on the rates and in depth interpretation of the causes of
the observed scalability curves. We will refer to this in future work.
Also, we would like to draw attention to the further advantage of
having a unified model applicable to systems of different fields that
allows the identification of similarities and analogies among,
apparently different, fields.


\begin{table*}
  \tiny
\centering
\begin{tabular}{rllllll}
param. & SQL server & wireless netw. & self-prop. part. & robot swarm & NAS BT & NAS SP \\ \hline 
$k_1$ & 0.1600614759 & 0.02889861707 & 1.390301954 & 1.475881283 & 0.03415882417 & 0.0987250043  \\ 
$k_2$ & 0.5640057846 & 0.1316419438 & 2.152415749 & 2.262085575 & 4.001963225 & 4.63175441  \\ 
$k_3$ & 0.3054490761 & 6.119334033 & 0 & 4.859955018  & 0 & 0.804438853\\ 
$k_4$ & 4.058864868 & 0.9700209090 & 8.499633576 & 9.341897861 & 8.899802172 & 7.93215055  \\ 
$k_5$ & 0.003989433297 & 0.003321168620 & 0 & 0.002558004976 & 0 & 3.73031451$\times 10^{-9}$ \\ 
$k_6$ & 0.5243111486 & 0.1970954618 & 0 & 0.3318341624  & 0 & 4.20617254\\ 
$k_7$ & 9.567349145 & 9.898300316 & 0 & 7.214769884 & 0 & 9.83304445 \\ 
$c_s$ & 0.212894142 & 0.05369458128 & 0.276059326 & 0.219971588 & 1.988764045 & 1.727748691 \\ 
MSE & $8.21545\times 10^{-4}$ & $1.21005\times 10^{-6}$ & $2.42738\times 10^{-4}$ & $2.77502\times 10^{-4}$ & $6.18738\times 10^{-3}$ & $4.37801\times 10^{-3}$ 
\end{tabular}
\caption{\label{tab:fitting}Fitted parameters for six different systems: multiprocessor SQL server, wireless network, self-propelled particles, robot swarm, NAS Parallel Benchmark Block Tridiagonal (BT) and scalar Pentadiagonal (SP); results are shown in Fig.~\ref{fig:fitting}. The row `MSE' indicates the mean squared error from the data points.}
\end{table*}





\section{Conclusion}\label{sec:discussion}

We have presented a new general model to study scalability of parallel
systems composed of multiple units (\textit{e.g.}, supercomputer composed of
CPUs, and artificial swarms composed of robots). Our model covers all
relevant scenarios of ideal concurrency, contention-limited systems,
and diminishing returns. In addition, our model is derived from a
microscopic description of state transitions of the units comprising the system. The units can be in three states that change over
time due to interactions with other units. The model can be
represented as a probabilistic state machine that describes the mean
behavior of a unit (\textit{microscopic} model). From the individual
state machine, we show how to derive the \textit{macroscopic}
population model that describes the scalability behavior of the whole
system. 
The derivation of the model from the mechanistic description of
interactions between the constituent units is one of the main
differences with previous scalability laws. Amdahl's law and
Gustafson's law are phenomenological models that are not derived from
first principles.  Gunther's USL is well motivated by queuing theory
but is still based on phenomenological assumptions, such as nonlinear
waiting times in the state-dependent server model. Our model has more
explanatory power by providing a direct link between system behavior and microscopic
behaviors of its units. However, increased explanatory power comes with
the cost of higher model complexity. The model has nine parameters: the seven transition rates that describe the frequency of state change
by the units, and the two contribution coefficients that describe the work contributed by a solitary and an interacting unit towards task
completion.
While the number of parameters is larger than in previous models, the
separation between transition rates~$k_i$ (mechanistic function) and
contribution coefficients~$c_s,c_g$ (task-specific performance) allows a better
understanding of the observed scalability behavior. For example, we interpret the cause of superlinear speedups as discussed
in Sec.~\ref{sec:superlinear}.

The transition rates~$k_i$ of a system can be determined in two alternative ways: by fitting the rates to system data (as shown in Sec.~\ref{sec:fitting}) or by observing and measuring individual transitions. 
While fitting parameters operates on data at the macroscopic level (system performance), measuring individual transitions represents a microscopic approach.
Individual transition rates can be obtained in
 systems that control the units using finite state machines, or behaviors, that may be mapped to our model's state machine. 
In a robot swarm, for example, robots can be assigned to state \textit{fermo} when their internal control state is \textit{robot-avoidance} due to physical proximity to other robots.
The problem of finding the correct model parameters is a special form of nonlinear system identification for gray-box models~\citep{nelles2013nonlinear}. 
Transition rates can potentially be estimated or even computed by either measuring the time spent by each unit in productive, collaborative, and unproductive states or by observing and counting individual state transitions. For instance, in a computing system, idle and busy times can be measured to estimate transition rates. A~better approach would be to count individual transitions while keeping track of populations in each state at the time of the observed transition.
Hence, complex machine behaviors can be simplified to observable state transitions between our simple three states (\textit{solo}, \textit{grupo}, \textit{fermo}), which allows to determine the model parameters. Based on our model and estimated rates~$k_i$, one can then predict the system behavior and study its potential for scalability.

We have focused on the asymptotic behavior (\textit{i.e.}, at
equilibrium) of the system, however the analysis can be extended to
the system's transient dynamics before reaching equilibrium. Such an analysis
could potentially provide additional information to improve the system
design. Similarly to previous models of scalability, our model is also
limited to static systems that do not vary the number of units
(\textit{e.g.}, in robotics see~\citep{buyya2009cloud,wahby2019collective,mayya2019voluntary,Rausch:SwInt:2019}) or
the task (\textit{e.g.},~\citep{mataric2003multi}) at runtime. Our model,
however, has potential to be extended for such cases by
modifying the state transitions to include `birth' and `death' of units
(system size variation), and by introducing time-varying parameters (dynamic tasks). Such features may be relevant in open systems where
units are added and removed online, for example, in on-demand cloud
computing which is based on dynamic use of
resources~\citep{buyya2009cloud}.

Our approach of starting from the description of changes of an individual
unit's state to derive a macroscopic model is common practice in
swarm
robotics~\citep{martinoli99,Reina:PLOSONE:2015,hamann2018book}
and collective behavior studies~\citep{Marshall:PONE:2019}. Such an
approach follows the line of thought of \textit{machine
behavior}~\citep{Rahwan2019} that advocates the necessity to
understand the cause of its behavior in order to fully control the machine. For example, the current efforts towards blackbox approaches in artificial
intelligence require a new
interdisciplinary approach in order to progress. In this study, we
show that through a collective behavior approach, we could contribute
to the understanding of parallel computing. Additionally, we
contribute towards a unified description of systems in diverse areas
such as parallel computing, sensor networks, and swarm robotics. In future work, we
plan to study whether our general model can also describe systems from other fields. In addition, other unified
models to describe engineering systems could be derived through
ethology and behavioral ecology methods, as it is commonly done in
swarm robotics~\citep{hamann2018book}.
Deeper insights may be obtained by studying scalability of systems with a set of interdisciplinary methods and with models of wide applicability.



\section*{Acknowledgments}

Both authors thank Neil J. Gunther and Gabriele Valentini for helpful comments about an earlier draft, remaining errors are ours. 
HH thanks the organizers and attendees of the Lakeside Research Days 2019 (Lakeside Labs and Alpen-Adria-Universität Klagenfurt) as well as Payam Zahadat for discussions and input. AR thanks Manuel L\'{o}pez-Ib\'{a}\~{n}ez for the help in identifying an appropriate fitting algorithm, and Arindam Saha for useful discussions on dynamical systems analysis. 
The authors thank the developers of many open-source/openly provided tools, such as
gnuplot,\footnote{
\url{http://www.gnuplot.info/}} emacs,\footnote{
\url{https://www.gnu.org/software/emacs/}} WebPlotDigitizer,\footnote{
\url{https://automeris.io/WebPlotDigitizer/}} and Overleaf.\footnote{
\url{https://www.overleaf.com/}} AR also acknowledges funding by the Belgian F.R.S.-FNRS of which he is Charg\'{e} de Recherches.


\section*{Appendix}
\section*{Stability analysis}
Our model, described as an ODE system in Eq.(12) of the main text, is derived from the seven transitions of Eqs.~(5)-(11) using van Kampen's expansion~\citep{vankampen}. Therefore, the terms $s$, $g$, and $f$ indicate unit concentrations $s=|S|/V$, $g=|G|/V$, and $f=|F|/V$, where $|S|$, $|G|$, and $|F|$ are the number of units in each of the three states \textit{solo}, \textit{grupo}, and \textit{fermo}, respectively, and the term $V$ is the system volume. As we are interested in studying the population dynamics for increasing concentrations of units for a fixed volume, for simplicity we fix $V=1$. Therefore, in our analysis, the concentrations $s$, $g$, and $f$ are equivalent to the population sizes $|S|$, $|G|$, and $|F|$.

In our model, the system size~$N$ is conserved, that is, the system units only change state but do not change in number (no birth or death); mathematically indicated by~$N=f+m+c$. Therefore, we can rewrite the ODE system of main text Eq.~(12) by substituting the subpopulation of units in state $G$ with~$g=N-s-f$. This substitution allows us to remove the second ODE. The resulting ODE system is 

\begin{equation}
    \left\{
    \begin{aligned}
      \frac{ds}{dt} &=-2k_1s^2-k_2s(N-s-f)-k_3sf+k_4(N-s-f) \\
      \frac{df}{dt} &= 2k_5 (N-s-f)^2 + k_6 f(N-s-f) - k_7 f \,
    \end{aligned}
    \right. .
    \label{eq:model2d}
\end{equation}
Writing the model in the form of Eq.~\eqref{eq:model2d} has two advantages compared to main text Eq.~12. Having a system of two equations, instead of three, simplifies its analysis. The ODE system now explicitly includes the system size, as parameter~$N$.

By varying transition rates $k_i$ (for $i \in \{1,\dots,7\}$), we show how our general model can flexibly approximate the dynamics of previous scalability laws, in particular of Gustafson's, Amdahl's, and Gunther's laws. Note that we assume that initially (at time $t=0$) the system has only \textit{solo} units: 
\begin{equation}
    s(0)=N\,, \qquad g(0)=0\,, \qquad f(0)=0\,.
    \label{eq:initialState}
\end{equation}

\subsection*{Gustafson's law: the ideal concurrency regime}

We set $k_2=k_3=k_5=k_6=k_7=0$, and we only keep $\{k_1,k_4\}>0$. In this way, Eq.~\eqref{eq:model2d} reduces to the single ODE
\begin{equation}
\label{eq:model-perm}
\frac{ds}{dt} =-2k_1s^2+k_4(N-s-f) \;,
\end{equation}
because $\frac{df}{dt}=0$. 
This equation has a single stable fixed point at
\begin{equation}
\label{eq:fp-perm}
s^*=\frac{\sqrt{k_4 (k_4 + 8 k_1 N)}-k_4}{4 k_1} \;,
\end{equation}
and we have $g^*=N-s^*$ and $f^*=0$. 


\subsection*{Amdahl's law: the contention-limited regime}

\begin{figure}
    \centering
    \begin{subfigure}[c]{0.7\textwidth}
      \includegraphics[width=1\textwidth]{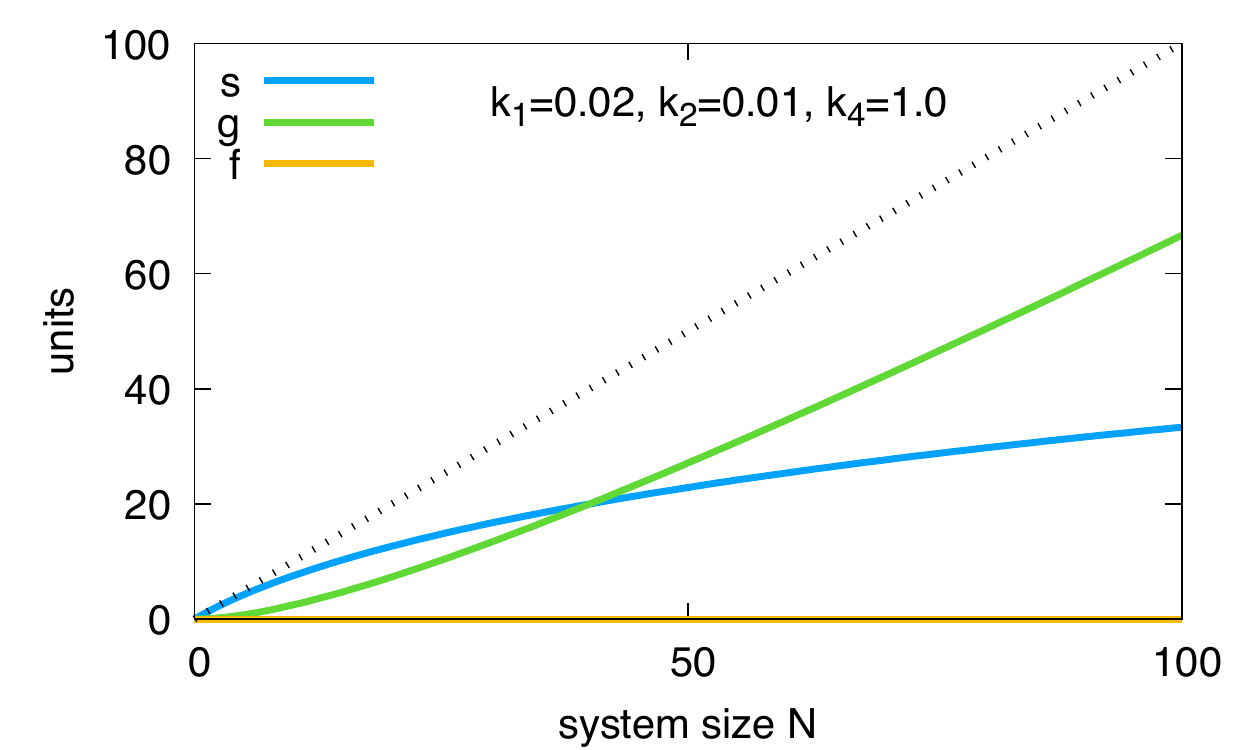}
      \subcaption{crowding rate $k_2=0.01=\frac{k_1}{2}$, $k_4=1.0$}
    \end{subfigure}
     \begin{subfigure}[c]{0.7\textwidth}
      \includegraphics[width=1\textwidth]{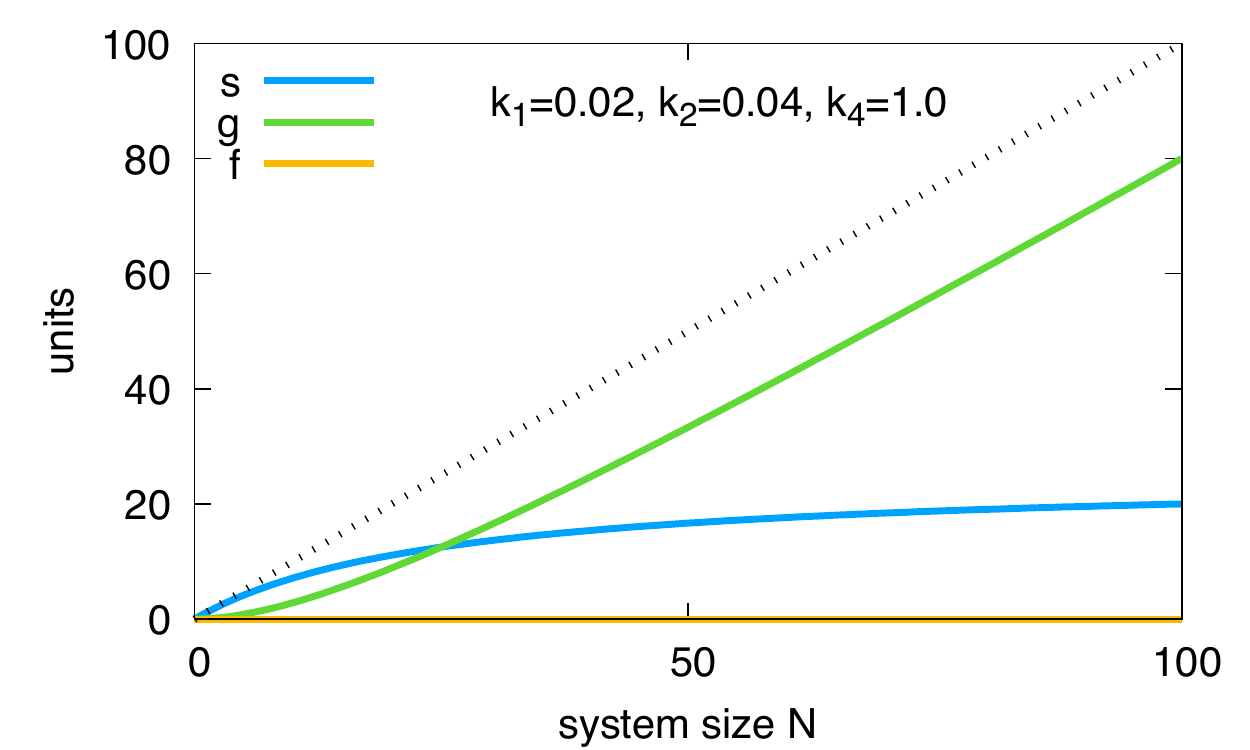}
      \subcaption{crowding rate $k_2=0.04=2k_1$, $k_4=1.0$}
    \end{subfigure}
    \begin{subfigure}[c]{0.7\textwidth}
      \includegraphics[width=1\textwidth]{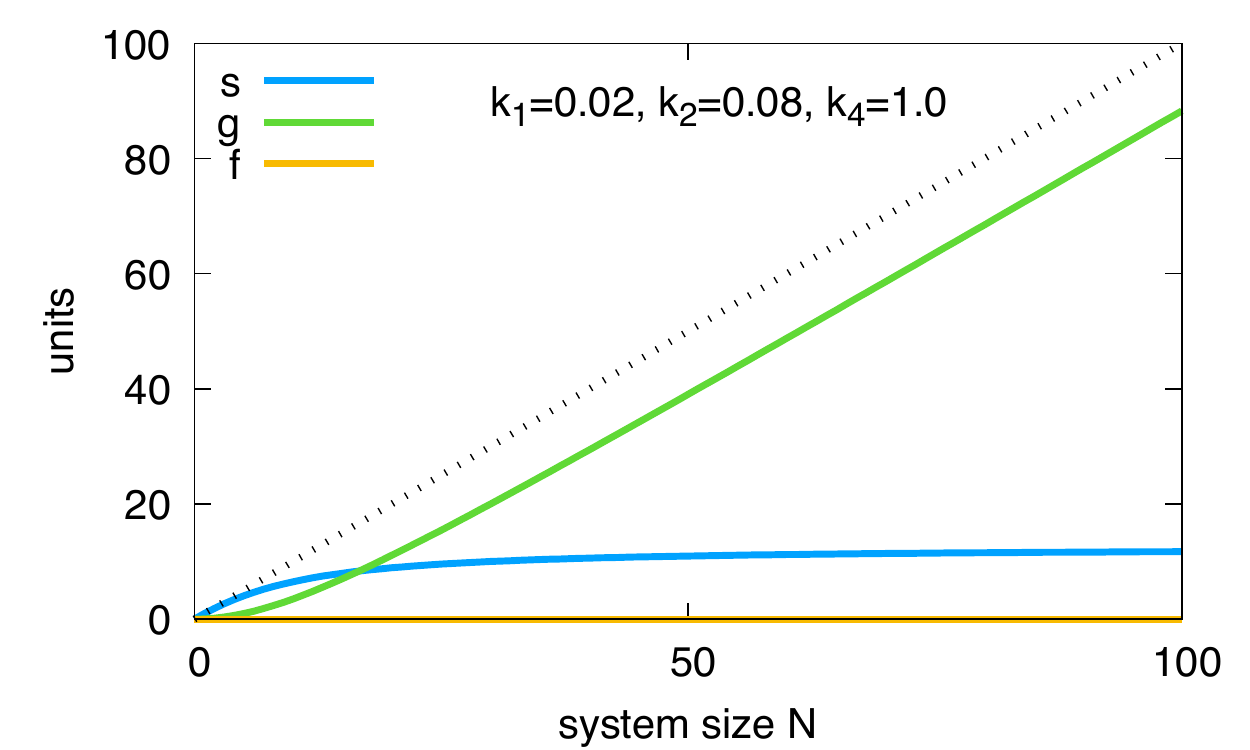}
      \subcaption{crowding rate $k_2=0.08=4k_1$, $k_4=1.0$}
    \end{subfigure}
    \caption{Amdahl's law: Increasing the positive feedback of crowding ($k_2$), in a system with simple interference ($k_1$) and decay ($k_4$), flattens the fraction of \textit{solo} units for increasing system size~$N$.}
    \label{fig:amdahl:supp}
\end{figure}

We set $k_3=k_5=k_6=k_7=0$, and we only keep $\{k_1,k_2,k_4\}>0$. The only difference from Gustafson's law is $k_2>0$ that corresponds to the transition $S+G\rightarrow 2G$ from main text Eq.~6.
As before, from the ODE system of Eq.~\eqref{eq:model2d} we obtain a single equation (because $\frac{df}{dt}=0$):
\begin{equation}
\label{eq:model-amdahl}
\frac{ds}{dt} =-2k_1s^2-k_2 s (N-s-f)+k_4(N-s-f) \;,
\end{equation}
which has a single stable fixed point at
\begin{equation}
\label{eq:fp-amdahl}
s^*=\frac{k_4 + k_2 N - \sqrt{k_4^2 + 8 k_1 k_4 N - 2 k_2 k_4 N + k_2^2 N^2}}{2 (-2 k_1 + k_2)} \;.
\end{equation}
This fixed point always exists, except for $k_2=2k_1$ due to a singularity and the fixed point of the system becomes
\begin{equation}
\label{eq:fp-amdahl-singularity}
s^*=\frac{k_4 N}{k_4+k_2 N} \;.
\end{equation}
As before, in both Eq.~\eqref{eq:fp-amdahl} and Eq.~\eqref{eq:fp-amdahl-singularity}, we have $g^*=N-s^*$ and $f^*=0$.
In this case, when the recruitment~$k_2$ is relatively high, the \textit{solo} fraction grows with system size~$N$ only for small values, smaller than~$N_c$. Beyond~$N_c$, it saturates to a value that decreases with $k_2/k_4$ (see Fig.~\ref{fig:amdahl:supp} for examples)

\subsection*{The diminishing returns regime}

\begin{figure}
    \centering
    \begin{subfigure}[c]{0.7\textwidth}
      \includegraphics[width=1\textwidth]{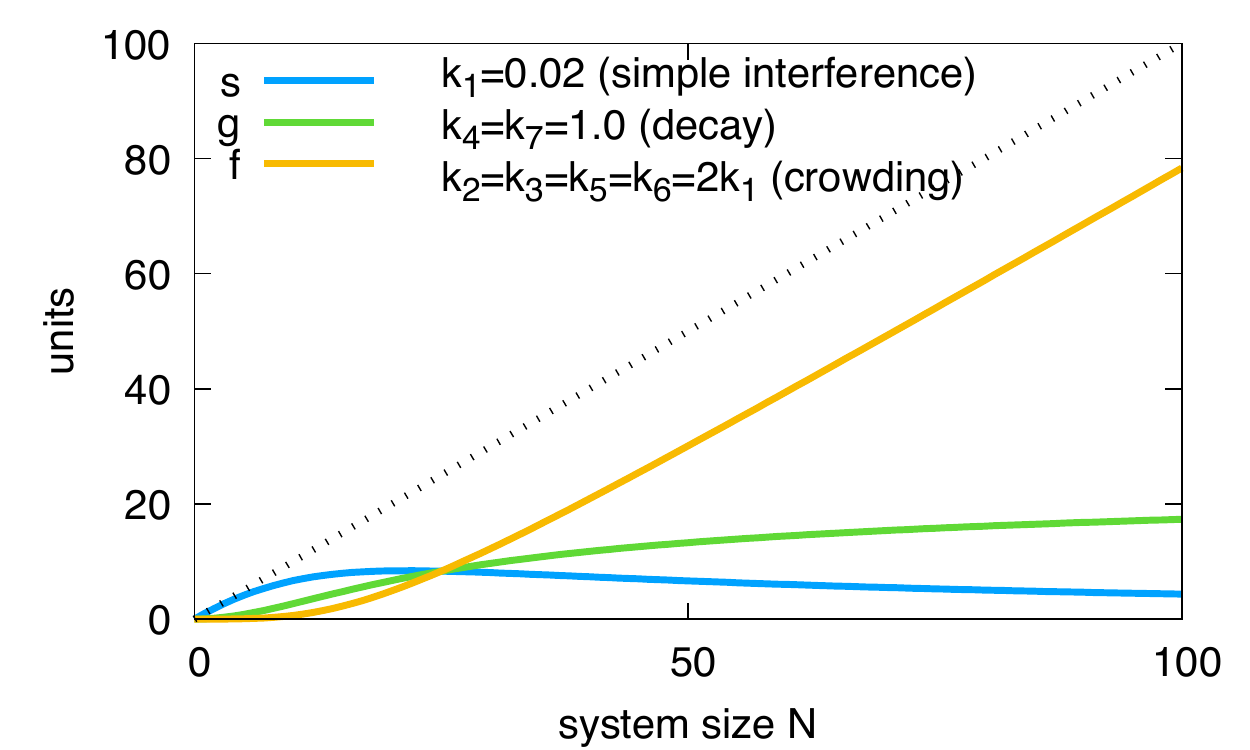}
      \subcaption{decay rate $k_4=k_7=1.0$}
    \end{subfigure}
     \begin{subfigure}[c]{0.7\textwidth}
      \includegraphics[width=1\textwidth]{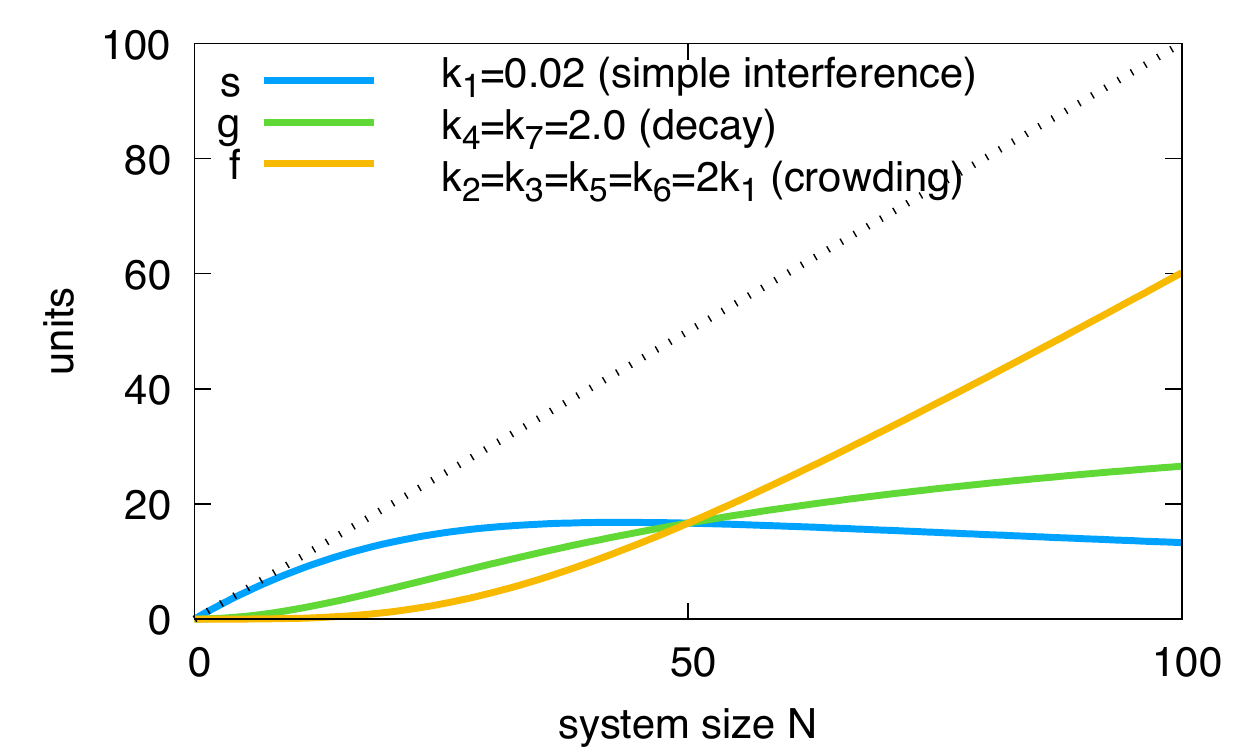}
      \subcaption{decay rate $k_4=k_7=2.0$}
    \end{subfigure}
    \begin{subfigure}[c]{0.7\textwidth}
      \includegraphics[width=1\textwidth]{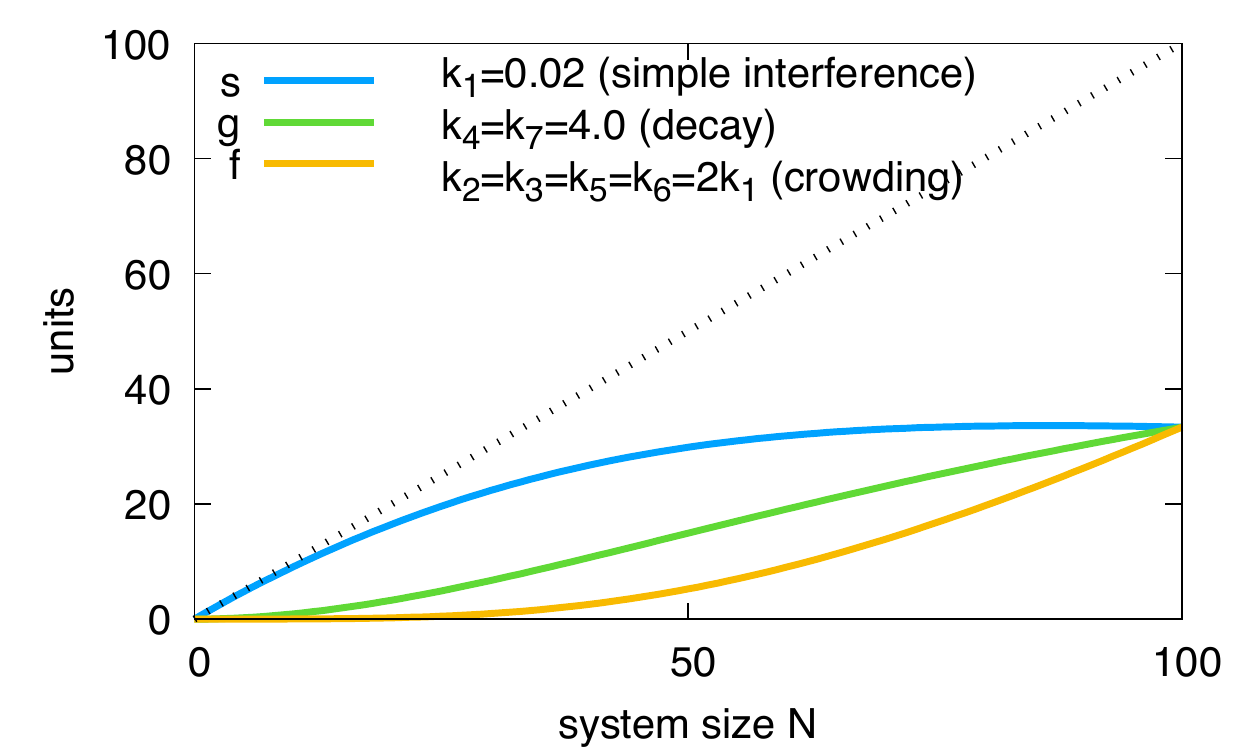}
      \subcaption{decay rate $k_4=k_7=4.0$\label{fig:case2k1_c}}
    \end{subfigure}
    \caption{Examples for diminishing returns. By increasing the frequency of the decay rates~$k_4$ and~$k_7$, an increasing fraction of units stabilizes in state \textit{solo}.}
    \label{fig:case2k1}
\end{figure}

\begin{figure}
    \centering
    \includegraphics[width=0.7\textwidth]{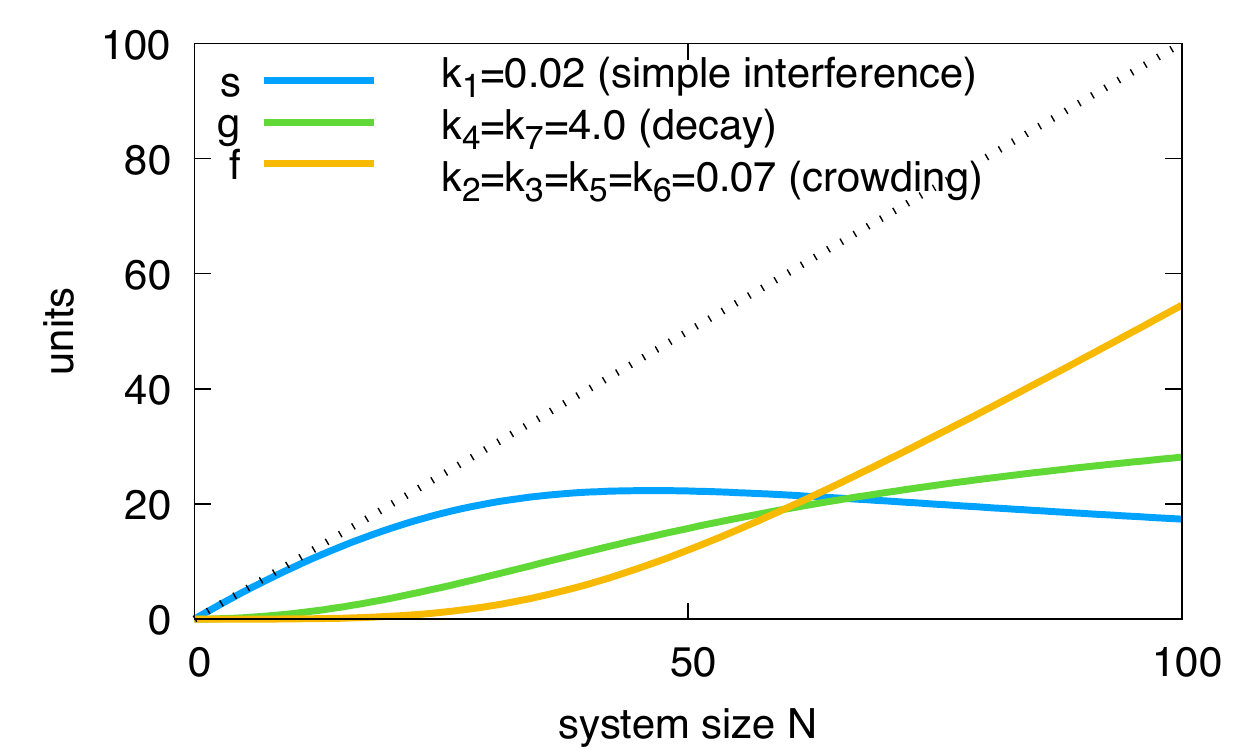}
    \caption{Example of diminishing return regime. By increasing system size~$N$ beyond $\approx 40$ units, the number of units in state \textit{solo} decreases. Assuming contribution coefficients~$c_s=1$ and~$c_g=0$, the system's performance would also decrease. Compared to Fig.~\ref{fig:case2k1_c}, a higher value of the `crowding' rates leads to an earlier and more pronounced decrease of~$s$.}
    \label{fig:full}
\end{figure}

We observe diminishing returns when the performance decreases with the system size $N$ for~$N > N_c$. If we assume performance coefficients $c_s=1$ and $c_g=0$, this type of behavior occurs when the number of units in the state \textit{solo} increases for small values of $N$ and collapses for high~$N$.
Interpreting the system's behavior through its dynamics (\textit{cf.} Figs.~\ref{fig:case2k1} and~\ref{fig:full}), we expect to observe diminishing returns when \textit{solo} units relatively slowly move to state \textit{grupo} (small rates~$k_1$ and~$k_2$), while they relatively quickly move to state \textit{fermo} (high rates~$k_5$ and~$k_6$). Mathematically, we can indicate that when~$k_5+k_6 > k_1+k_2$ we observe the diminishing results regime. Therefore, we study the system with the following parameters:
\begin{equation}
    k_2=k_3=k_5=k_6=2k_1, \qquad k_4=k_7 \;,
    \label{eq:simplification2k1}
\end{equation}
which means that the crowding rate is double the interaction rate between \textit{solo} units~($k_1$). Additionally, we simplify the analysis by imposing the same decay from states \textit{grupo} and \textit{fermo}. Using the parameters of Eq.~\eqref{eq:simplification2k1}, the stable fixed points of the ODE system of Eq.~\eqref{eq:model2d} is
\begin{equation}
\begin{aligned}
    s^*&=\frac{2 k_4^2 N}{4 k_1^2 N^2+\sqrt{16 k_1^4 N^4+48 k_4 k_1^3 N^3-4 k_4^2 k_1^2 N^2+4 k_4^3 k_1 N+k_4^4}+2 k_4 k_1 N+k_4^2}\\
    f^*&=-\frac{24 k_1^3 N^3-\left(2 k_1 N+k_4\right) \sqrt{16 k_1^4 N^4+48 k_4 k_1^3 N^3-4 k_4^2 k_1^2 N^2+4 k_4^3 k_1 N+k_4^4}+4
   k_4^2 k_1 N+k_4^3}{8 k_1^2 k_4 N-16 k_1^3 N^2}\\
   g*&=N-s^*-f^* \;.
\end{aligned}
\label{eq:fp-case2k1}
\end{equation}
Fig.~\ref{fig:case2k1} shows three examples of the dynamics of Eq.~\eqref{eq:fp-case2k1}. The reduction in \textit{solo} units is more pronounced for higher crowding rates, as illustrated in Fig.~\ref{fig:full}. 

\paragraph*{Approximation} For the above diminishing returns scenario (Eq.~\ref{eq:fp-case2k1}), we can approximate 
\begin{align}
&16 k_1^4 N^4+48 k_4 k_1^3 N^3-4 k_4^2 k_1^2 N^2+4 k_4^3 k_1 N+k_4^4\notag\\
&\approx
\left(4 k_1^2 N^2+2 k_4 k_1 N+k_4^2\right){}^2 \;.
\end{align}
In this way, the first equation of Eq.~\eqref{eq:fp-case2k1} reduces to
\begin{equation}
    s^* = \frac{N}{\frac{k_2^2}{k_4^2}N^2+\frac{k_2}{k_4}N+1} \;,
    \label{eq:simpleUSL}
\end{equation}
which closely corresponds to the Gunther's USL (main text Eq.~3) by considering $\kappa=\sigma^2$ and $\sigma=\frac{k_2}{k_4}$.

\pagebreak

\footnotesize
\bibliographystyle{abbrvnat}
\bibliography{./main}

\begin{thebibliography}{61}
\providecommand{\natexlab}[1]{#1}
\providecommand{\url}[1]{\texttt{#1}}
\expandafter\ifx\csname urlstyle\endcsname\relax
  \providecommand{\doi}[1]{doi: #1}\else
  \providecommand{\doi}{doi: \begingroup \urlstyle{rm}\Url}\fi

\bibitem[Allen(1990)]{Allen90}
A.~O. Allen.
\newblock \emph{Probability, Statistics, and Queueing Theory with Computer
  Science Applications}.
\newblock Academic Press Professional, Inc., USA, 1990.
\newblock ISBN 0120510510.

\bibitem[Amdahl(1967)]{amdahl1967validity}
G.~M. Amdahl.
\newblock Validity of the single processor approach to achieving large scale
  computing capabilities.
\newblock In \emph{AFIPS Conference Proceedings}, pages 483--485. ACM, 1967.

\bibitem[Archibald and Baer(1986)]{archibald1986cache}
J.~Archibald and J.-L. Baer.
\newblock Cache coherence protocols: Evaluation using a multiprocessor
  simulation model.
\newblock \emph{ACM Transactions on Computer Systems (TOCS)}, 4\penalty0
  (4):\penalty0 273--298, 1986.

\bibitem[{ASF - Apache Software Foundation}()]{ApacheHadoop}
{ASF - Apache Software Foundation}.
\newblock Apache hadoop.
\newblock \url{http://hadoop.apache.org}.

\bibitem[Ballard et~al.(2011)Ballard, Demmel, Holtz, and
  Schwartz]{ballard2011minimizing}
G.~Ballard, J.~Demmel, O.~Holtz, and O.~Schwartz.
\newblock Minimizing communication in numerical linear algebra.
\newblock \emph{SIAM Journal on Matrix Analysis and Applications}, 32\penalty0
  (3):\penalty0 866--901, 2011.

\bibitem[Berman et~al.(2011)Berman, Lindsey, Sakar, Kumar, and
  Pratt]{berman11b}
S.~Berman, Q.~Lindsey, M.~S. Sakar, V.~Kumar, and S.~C. Pratt.
\newblock Experimental study and modeling of group retrieval in ants as an
  approach to collective transport in swarm robotic systems.
\newblock \emph{Proceedings of the IEEE}, 99\penalty0 (9):\penalty0 1470--1481,
  Sept 2011.
\newblock ISSN 0018-9219.
\newblock \doi{10.1109/JPROC.2011.2111450}.

\bibitem[Bettstetter(2004)]{bettstetter04}
C.~Bettstetter.
\newblock On the connectivity of {A}d {H}oc networks.
\newblock \emph{The Computer Journal}, 47\penalty0 (4):\penalty0 432--447,
  2004.

\bibitem[Bjerknes and Winfield(2013)]{Bjerknes2013}
J.~D. Bjerknes and A.~F.~T. Winfield.
\newblock On fault tolerance and scalability of swarm robotic systems.
\newblock In M.~{Ani Hsieh} and G.~Chirikjian, editors, \emph{Distributed
  Autonomous Robotic Systems: The 10th International Symposium}, volume 104 of
  \emph{Springer Tracts in Advanced Robotics}, pages 431--444. Springer Berlin
  Heidelberg, Berlin, Heidelberg, 2013.
\newblock \doi{10.1007/978-3-642-32723-0_31}.

\bibitem[Bjerknes et~al.(2007)Bjerknes, Winfield, and Melhuish]{bjerknes07}
J.~D. Bjerknes, A.~Winfield, and C.~Melhuish.
\newblock An analysis of emergent taxis in a wireless connected swarm of mobile
  robots.
\newblock In Y.~Shi and M.~Dorigo, editors, \emph{IEEE Swarm Intelligence
  Symposium}, pages 45--52, Los Alamitos, CA, 2007. IEEE Press.

\bibitem[Bogdan and Marculescu(2009)]{bogdan09}
P.~Bogdan and R.~Marculescu.
\newblock Statistical physics approaches for network-on-chip traffic
  characterization.
\newblock In \emph{Proceedings of the 7th IEEE/ACM International Conference on
  Hardware/Software Codesign and System Synthesis}, CODES+ISSS '09, page
  461–470, New York, NY, USA, 2009. Association for Computing Machinery.
\newblock ISBN 9781605586281.
\newblock \doi{10.1145/1629435.1629498}.
\newblock URL \url{https://doi.org/10.1145/1629435.1629498}.

\bibitem[Bogdan and Marculescu(2011)]{Bogdan2011}
P.~Bogdan and R.~Marculescu.
\newblock Non-stationary traffic analysis and its implications on multicore
  platform design.
\newblock \emph{IEEE Transactions on Computer-Aided Design of Integrated
  Circuits and Systems}, 30\penalty0 (4):\penalty0 508--519, apr 2011.
\newblock ISSN 0278-0070.
\newblock \doi{10.1109/TCAD.2011.2111270}.
\newblock URL \url{http://ieeexplore.ieee.org/document/5737846/}.

\bibitem[Buyya et~al.(2009)Buyya, Yeo, Venugopal, Broberg, and
  Brandic]{buyya2009cloud}
R.~Buyya, C.~S. Yeo, S.~Venugopal, J.~Broberg, and I.~Brandic.
\newblock Cloud computing and emerging it platforms: Vision, hype, and reality
  for delivering computing as the 5th utility.
\newblock \emph{Future Generation computer systems}, 25\penalty0 (6):\penalty0
  599--616, 2009.

\bibitem[Dumitras and Marculescu(2003)]{dumitras03}
T.~Dumitras and R.~Marculescu.
\newblock On-chip stochastic communication [soc applications].
\newblock In \emph{2003 Design, Automation and Test in Europe Conference and
  Exhibition}, pages 790--795. IEEE, 2003.

\bibitem[Espenson(1995)]{espenson95}
J.~H. Espenson.
\newblock \emph{Chemical kinetics and reaction mechanisms}, volume 102.
\newblock McGraw-Hill, 1995.

\bibitem[Farkas et~al.(1995)Farkas, Vranesic, and Stumm]{farkas1995scalable}
K.~Farkas, Z.~Vranesic, and M.~Stumm.
\newblock Scalable cache consistency for hierarchically structured
  multiprocessors.
\newblock \emph{The Journal of Supercomputing}, 8\penalty0 (4):\penalty0
  345--369, 1995.

\bibitem[Feng et~al.(2020)Feng, Deng, Chen, Perc, and Kurths]{feng20}
M.~Feng, L.-J. Deng, F.~Chen, M.~Perc, and J.~Kurths.
\newblock The accumulative law and its probability model: an extension of the
  pareto distribution and the log-normal distribution.
\newblock \emph{Proceedings of the Royal Society A: Mathematical, Physical and
  Engineering Sciences}, 476\penalty0 (2237):\penalty0 20200019, 2020.
\newblock \doi{10.1098/rspa.2020.0019}.
\newblock URL
  \url{https://royalsocietypublishing.org/doi/abs/10.1098/rspa.2020.0019}.

\bibitem[Garnier et~al.(2007)Garnier, Gautrais, and Theraulaz]{garnier07}
S.~Garnier, J.~Gautrais, and G.~Theraulaz.
\newblock The biological principles of swarm intelligence.
\newblock \emph{Swarm Intelligence}, 1:\penalty0 3--31, 2007.
\newblock URL \url{http://dx.doi.org/10.1007/s11721-007-0004-y}.

\bibitem[Gillespie et~al.(2013)Gillespie, Hellander, and
  Petzold]{Gillespie2013}
D.~T. Gillespie, A.~Hellander, and L.~R. Petzold.
\newblock Perspective: Stochastic algorithms for chemical kinetics.
\newblock \emph{The Journal of Chemical Physics}, 138\penalty0 (17):\penalty0
  170901, 2013.
\newblock \doi{10.1063/1.4801941}.

\bibitem[Gittins et~al.(1989)Gittins, Glazebrook, and Weber]{gittins1989}
J.~Gittins, K.~Glazebrook, and R.~Weber.
\newblock \emph{Multi-armed bandit allocation indices}.
\newblock John Wiley \& Sons, 1989.

\bibitem[Gokturk et~al.(2008)Gokturk, Ercetin, and Gurbuz]{gokturk08}
M.~S. Gokturk, O.~Ercetin, and O.~Gurbuz.
\newblock Throughput analysis of {ALOHA} with cooperative diversity.
\newblock \emph{IEEE Communications Letters}, 12\penalty0 (6):\penalty0
  468--470, 2008.

\bibitem[Grama et~al.(2003)Grama, Kumar, Gupta, and
  Karypis]{grama2003introduction}
A.~Grama, V.~Kumar, A.~Gupta, and G.~Karypis.
\newblock \emph{Introduction to parallel computing}.
\newblock Pearson Education, 2003.

\bibitem[Gunther(1993)]{gunther93}
N.~J. Gunther.
\newblock A simple capacity model of massively parallel transaction systems.
\newblock In \emph{CMG National Conf.}, pages 1035--1044, 1993.

\bibitem[Gunther(2007)]{gunther07}
N.~J. Gunther.
\newblock \emph{Guerrilla Capacity Planning}.
\newblock Springer, 2007.

\bibitem[Gunther(2008)]{gunther2008general}
N.~J. Gunther.
\newblock A general theory of computational scalability based on rational
  functions.
\newblock \emph{arXiv preprint arXiv:0808.1431}, 2008.

\bibitem[Gunther et~al.(2015{\natexlab{a}})Gunther, Puglia, and
  Tomasette]{gunther15}
N.~J. Gunther, P.~Puglia, and K.~Tomasette.
\newblock Hadoop super-linear scalability: The perpetual motion of parallel
  performance.
\newblock \emph{ACM Queue}, 13\penalty0 (5), 2015{\natexlab{a}}.

\bibitem[Gunther et~al.(2015{\natexlab{b}})Gunther, Puglia, and
  Tomasette]{gunther15b}
N.~J. Gunther, P.~Puglia, and K.~Tomasette.
\newblock Hadoop super-linear scalability.
\newblock \emph{Commun. ACM}, 58\penalty0 (4):\penalty0 46--55,
  2015{\natexlab{b}}.

\bibitem[Gustafson(1988)]{gustafson1988}
J.~L. Gustafson.
\newblock Reevaluating {A}mdahl's law.
\newblock \emph{Commun. ACM}, 31\penalty0 (5):\penalty0 532--533, May 1988.
\newblock ISSN 0001-0782.
\newblock \doi{10.1145/42411.42415}.
\newblock URL \url{http://doi.acm.org/10.1145/42411.42415}.

\bibitem[Gustafson(1990)]{gustafson90}
J.~L. Gustafson.
\newblock Fixed time, tiered memory, and superlinear speedup.
\newblock In \emph{Proceedings of the Fifth Distributed Memory Computing
  Conference (DMCC5)}, pages 1255--1260, 1990.

\bibitem[Hamann(2013)]{hamann13a}
H.~Hamann.
\newblock Towards swarm calculus: Urn models of collective decisions and
  universal properties of swarm performance.
\newblock \emph{Swarm Intelligence}, 7\penalty0 (2-3):\penalty0 145--172, 2013.
\newblock URL \url{http://dx.doi.org/10.1007/s11721-013-0080-0}.

\bibitem[Hamann(2018{\natexlab{a}})]{hamann18b}
H.~Hamann.
\newblock Superlinear scalability in parallel computing and multi-robot
  systems: Shared resources, collaboration, and network topology.
\newblock In M.~Berekovic, R.~Buchty, H.~Hamann, D.~Koch, and T.~Pionteck,
  editors, \emph{Architecture of Computing Systems -- ARCS 2018}, pages 31--42,
  Cham, 2018{\natexlab{a}}. Springer International Publishing.
\newblock ISBN 978-3-319-77610-1.

\bibitem[Hamann(2018{\natexlab{b}})]{hamann2018book}
H.~Hamann.
\newblock \emph{Swarm Robotics: A Formal Approach}.
\newblock Springer, Cham, 2018{\natexlab{b}}.
\newblock \doi{10.1007/978-3-319-74528-2}.

\bibitem[Helmbold and McDowell(1990)]{helmbold90}
D.~P. Helmbold and C.~E. McDowell.
\newblock Modelling speedup (n) greater than n.
\newblock \emph{IEEE Transactions on Parallel and Distributed Systems},
  1\penalty0 (2):\penalty0 250--256, 1990.

\bibitem[Ijspeert et~al.(2001)Ijspeert, Martinoli, Billard, and
  Gambardella]{ijspeert01b}
A.~J. Ijspeert, A.~Martinoli, A.~Billard, and L.~M. Gambardella.
\newblock Collaboration through the exploitation of local interactions in
  autonomous collective robotics: The stick pulling experiment.
\newblock \emph{Autonomous Robots}, 11:\penalty0 149--171, 2001.
\newblock ISSN 0929-5593.
\newblock \doi{10.1023/A:1011227210047}.
\newblock URL \url{http://dx.doi.org/10.1023/A%3A1011227210047}.

\bibitem[Jin et~al.(1999)Jin, Frumkin, and Yan]{jin99}
H.~Jin, M.~Frumkin, and J.~Yan.
\newblock The {OpenMP} implementation of {NAS} parallel benchmarks and its
  performance.
\newblock Technical Report NAS-99-011, NASA Ames Research Center, 1999.

\bibitem[Khaluf et~al.(2017)Khaluf, Pinciroli, Valentini, and Hamann]{khaluf17}
Y.~Khaluf, C.~Pinciroli, G.~Valentini, and H.~Hamann.
\newblock The impact of agent density on scalability in collective systems:
  noise-induced versus majority-based bistability.
\newblock \emph{Swarm Intelligence}, 11\penalty0 (2):\penalty0 155--179, Jun
  2017.
\newblock ISSN 1935-3820.
\newblock \doi{10.1007/s11721-017-0137-6}.
\newblock URL \url{https://doi.org/10.1007/s11721-017-0137-6}.

\bibitem[Kirk and Wen-Mei(2016)]{kirk2016programming}
D.~B. Kirk and W.~H. Wen-Mei.
\newblock \emph{Programming massively parallel processors: a hands-on
  approach}.
\newblock Morgan kaufmann, 2016.

\bibitem[Lazer and Friedman(2007)]{lazer07}
D.~Lazer and A.~Friedman.
\newblock The network structure of exploration and exploitation.
\newblock \emph{Administrative Science Quarterly}, 52:\penalty0 667--694, 2007.

\bibitem[Lerman and Galstyan(2001)]{lerman2001general}
K.~Lerman and A.~Galstyan.
\newblock A general methodology for mathematical analysis of multi-agent
  systems.
\newblock \emph{ISI-TR-529, USC Information Sciences Institute, Marina del Rey,
  CA}, 2001.

\bibitem[Mahnke and Kaupu{\v{z}}s(1999)]{mahnke1999stochastic}
R.~Mahnke and J.~Kaupu{\v{z}}s.
\newblock Stochastic theory of freeway traffic.
\newblock \emph{Physical Review E}, 59\penalty0 (1):\penalty0 117, 1999.

\bibitem[Marshall et~al.(2019)Marshall, Reina, and Bose]{Marshall:PONE:2019}
J.~A.~R. Marshall, A.~Reina, and T.~Bose.
\newblock Multiscale modelling tool: Mathematical modelling of collective
  behaviour without the maths.
\newblock \emph{PLoS ONE}, 14\penalty0 (9):\penalty0 e0222906, 2019.
\newblock \doi{10.1371/journal.pone.0222906}.

\bibitem[Martinoli(1999)]{martinoli99}
A.~Martinoli.
\newblock \emph{Swarm Intelligence in Autonomous Collective Robotics: From
  Tools to the Analysis and Synthesis of Distributed Control Strategies}.
\newblock PhD thesis, Ecole Polytechnique F\'ed\'erale de Lausanne, 1999.

\bibitem[Martinoli et~al.(2004)Martinoli, Easton, and Agassounon]{martinoli04}
A.~Martinoli, K.~Easton, and W.~Agassounon.
\newblock Modeling swarm robotic systems: {A} case study in collaborative
  distributed manipulation.
\newblock \emph{Int. Journal of Robotics Research}, 23\penalty0 (4):\penalty0
  415--436, 2004.

\bibitem[Matari{\'c} et~al.(2003)Matari{\'c}, Sukhatme, and
  {\O}stergaard]{mataric2003multi}
M.~J. Matari{\'c}, G.~S. Sukhatme, and E.~H. {\O}stergaard.
\newblock Multi-robot task allocation in uncertain environments.
\newblock \emph{Autonomous Robots}, 14\penalty0 (2-3):\penalty0 255--263, 2003.

\bibitem[Mayya et~al.(2019)Mayya, Pierpaoli, and Egerstedt]{mayya2019voluntary}
S.~Mayya, P.~Pierpaoli, and M.~Egerstedt.
\newblock Voluntary retreat for decentralized interference reduction in robot
  swarms.
\newblock In \emph{Int. Conf. on Robotics and Automation (ICRA)}, pages
  9667--9673. IEEE, 2019.

\bibitem[Nelles(2013)]{nelles2013nonlinear}
O.~Nelles.
\newblock \emph{Nonlinear system identification: from classical approaches to
  neural networks and fuzzy models}.
\newblock Springer Science \& Business Media, 2013.

\bibitem[O'Grady et~al.(2007)O'Grady, Gross, Christensen, Mondada, Bonani, and
  Dorigo]{ogrady07}
R.~O'Grady, R.~Gross, A.~L. Christensen, F.~Mondada, M.~Bonani, and M.~Dorigo.
\newblock Performance benefits of self-assembly in a swarm-bot.
\newblock In \emph{IEEE/RSJ International Conference on Intelligent Robots and
  Systems (IROS)}, pages 2381--2387, Oct 2007.
\newblock \doi{10.1109/IROS.2007.4399424}.

\bibitem[Rahwan et~al.(2019)Rahwan, Cebrian, Obradovich, Bongard, Bonnefon,
  Breazeal, Crandall, Christakis, Couzin, Jackson, Jennings, Kamar, Kloumann,
  Larochelle, Lazer, McElreath, Mislove, Parkes, Pentland, Roberts, Shariff,
  Tenenbaum, and Wellman]{Rahwan2019}
I.~Rahwan, M.~Cebrian, N.~Obradovich, J.~Bongard, J.-F. Bonnefon, C.~Breazeal,
  J.~W. Crandall, N.~A. Christakis, I.~D. Couzin, M.~O. Jackson, N.~R.
  Jennings, E.~Kamar, I.~M. Kloumann, H.~Larochelle, D.~Lazer, R.~McElreath,
  A.~Mislove, D.~C. Parkes, A.~S. Pentland, M.~E. Roberts, A.~Shariff, J.~B.
  Tenenbaum, and M.~Wellman.
\newblock Machine behaviour.
\newblock \emph{Nature}, 568\penalty0 (7753):\penalty0 477--486, 2019.
\newblock \doi{10.1038/s41586-019-1138-y}.

\bibitem[Rausch et~al.(2019)Rausch, Reina, Simoens, and
  Khaluf]{Rausch:SwInt:2019}
I.~Rausch, A.~Reina, P.~Simoens, and Y.~Khaluf.
\newblock Coherent collective behaviour emerging from decentralised balancing
  of social feedback and noise.
\newblock \emph{Swarm Intelligence}, 13\penalty0 (3--4):\penalty0 321--345,
  2019.
\newblock \doi{10.1007/s11721-019-00173-y}.

\bibitem[Reina et~al.(2015)Reina, Valentini, Fern\'{a}ndez-Oto, Dorigo, and
  Trianni]{Reina:PLOSONE:2015}
A.~Reina, G.~Valentini, C.~Fern\'{a}ndez-Oto, M.~Dorigo, and V.~Trianni.
\newblock A design pattern for decentralised decision making.
\newblock \emph{PLoS ONE}, 10\penalty0 (10):\penalty0 e0140950, 2015.
\newblock \doi{10.1371/journal.pone.0140950}.

\bibitem[Resnick(1994)]{resnick94}
M.~Resnick.
\newblock \emph{Turtles, Termites, and Traffic Jams}.
\newblock MIT Press, 1994.

\bibitem[Ribeiro et~al.(2012)Ribeiro, Castro, Marangonzova-Martin, M{\'e}haut,
  de~Freitas, and da~Silva~Martins]{ribeiro12}
C.~P. Ribeiro, M.~Castro, V.~Marangonzova-Martin, J.-F. M{\'e}haut, H.~C.
  de~Freitas, and C.~A.~P. da~Silva~Martins.
\newblock Evaluating {CPU} and memory affinity for numerical scientific
  multithreaded benchmarks on multi-cores.
\newblock \emph{IADIS International Journal on Computer Science and Information
  Systems (IJCSIS)}, 7:\penalty0 79--93, 2012.

\bibitem[Roberts(1975)]{roberts75}
L.~G. Roberts.
\newblock {ALOHA} packet system with and without slots and capture.
\newblock \emph{SIGCOMM Comput. Commun. Rev.}, 5\penalty0 (2):\penalty0
  28–42, Apr. 1975.
\newblock ISSN 0146-4833.
\newblock \doi{10.1145/1024916.1024920}.
\newblock URL \url{https://doi.org/10.1145/1024916.1024920}.

\bibitem[Rosenfeld et~al.(2006)Rosenfeld, Kaminka, and Kraus]{rosenfeld06}
A.~Rosenfeld, G.~A. Kaminka, and S.~Kraus.
\newblock A study of scalability properties in robotic teams.
\newblock In P.~Scerri, R.~Vincent, and R.~Mailler, editors, \emph{Coordination
  of Large-Scale Multiagent Systems}, pages 27--51. Springer US, Boston, MA,
  2006.
\newblock ISBN 978-0-387-27972-5.
\newblock \doi{10.1007/0-387-27972-5_2}.
\newblock URL \url{https://doi.org/10.1007/0-387-27972-5_2}.

\bibitem[Sayama et~al.(2013)Sayama, Pestov, Schmidt, Bush, Wong, Yamanoi, and
  Gross]{sayama13}
H.~Sayama, I.~Pestov, J.~Schmidt, B.~J. Bush, C.~Wong, J.~Yamanoi, and
  T.~Gross.
\newblock Modeling complex systems with adaptive networks.
\newblock \emph{Computers \& Mathematics with Applications}, 65\penalty0
  (10):\penalty0 1645--1664, 2013.
\newblock ISSN 0898-1221.
\newblock \doi{https://doi.org/10.1016/j.camwa.2012.12.005}.
\newblock URL
  \url{https://www.sciencedirect.com/science/article/pii/S0898122112007018}.
\newblock Grasping Complexity.

\bibitem[Shang and Bouffanais(2014)]{shang14}
Y.~Shang and R.~Bouffanais.
\newblock Influence of the number of topologically interacting neighbors on
  swarm dynamics.
\newblock \emph{Scientific Reports}, 4\penalty0 (4184), 2014.

\bibitem[Storn and Price(1997)]{storn1997differential}
R.~Storn and K.~Price.
\newblock Differential evolution--a simple and efficient heuristic for global
  optimization over continuous spaces.
\newblock \emph{Journal of Global Optimization}, 11\penalty0 (4):\penalty0
  341--359, 1997.

\bibitem[Talamali et~al.(2020)Talamali, Bose, Haire, Xu, Marshall, and
  Reina]{Talamali:SwInt:2019}
M.~S. Talamali, T.~Bose, M.~Haire, X.~Xu, J.~A.~R. Marshall, and A.~Reina.
\newblock Sophisticated collective foraging with minimalist agents: A swarm
  robotics test.
\newblock \emph{Swarm Intelligence}, 14\penalty0 (1):\penalty0 25--56, 2020.
\newblock \doi{10.1007/s11721-019-00176-9}.

\bibitem[{van Kampen}(1981)]{vankampen}
N.~G. {van Kampen}.
\newblock \emph{Stochastic Processes in Physics and Chemistry}.
\newblock North-Holland, Amsterdam, 1981.

\bibitem[{Virtanen} et~al.(2020){Virtanen}, {Gommers}, {Oliphant}, {Haberland},
  {Reddy}, {Cournapeau}, {Burovski}, {Peterson}, {Weckesser}, {Bright}, {van
  der Walt}, {Brett}, {Wilson}, {Jarrod Millman}, {Mayorov}, {Nelson}, {Jones},
  {Kern}, {Larson}, {Carey}, {Polat}, {Feng}, {Moore}, {Vand erPlas},
  {Laxalde}, {Perktold}, {Cimrman}, {Henriksen}, {Quintero}, {Harris},
  {Archibald}, {Ribeiro}, {Pedregosa}, {van Mulbregt}, and
  {Contributors}]{SciPy}
P.~{Virtanen}, R.~{Gommers}, T.~E. {Oliphant}, M.~{Haberland}, T.~{Reddy},
  D.~{Cournapeau}, E.~{Burovski}, P.~{Peterson}, W.~{Weckesser}, J.~{Bright},
  S.~J. {van der Walt}, M.~{Brett}, J.~{Wilson}, K.~{Jarrod Millman},
  N.~{Mayorov}, A.~R.~J. {Nelson}, E.~{Jones}, R.~{Kern}, E.~{Larson},
  C.~{Carey}, {\.I}.~{Polat}, Y.~{Feng}, E.~W. {Moore}, J.~{Vand erPlas},
  D.~{Laxalde}, J.~{Perktold}, R.~{Cimrman}, I.~{Henriksen}, E.~A. {Quintero},
  C.~R. {Harris}, A.~M. {Archibald}, A.~H. {Ribeiro}, F.~{Pedregosa}, P.~{van
  Mulbregt}, and S.~.~. {Contributors}.
\newblock {SciPy 1.0: Fundamental Algorithms for Scientific Computing in
  Python}.
\newblock \emph{Nature Methods}, 17:\penalty0 261--272, 2020.
\newblock \doi{https://doi.org/10.1038/s41592-019-0686-2}.

\bibitem[Wahby et~al.(2019)Wahby, Petzold, Eschke, Schmickl, and
  Hamann]{wahby2019collective}
M.~Wahby, J.~Petzold, C.~Eschke, T.~Schmickl, and H.~Hamann.
\newblock Collective change detection: Adaptivity to dynamic swarm densities
  and light conditions in robot swarms.
\newblock In \emph{Artificial life: A hybrid of the European conference on
  artificial life (ECAL) and the international conference on the synthesis and
  simulation of living systems (ALIFE)}, pages 642--649. MIT Press, 2019.

\bibitem[Zhang et~al.(2007)Zhang, Neglia, Kurose, and
  Towsley]{zhang2007performance}
X.~Zhang, G.~Neglia, J.~Kurose, and D.~Towsley.
\newblock Performance modeling of epidemic routing.
\newblock \emph{Computer Networks}, 51\penalty0 (10):\penalty0 2867--2891,
  2007.

\end{thebibliography}



\end{document}